\documentclass[english]{article}
\usepackage[T1]{fontenc}
\usepackage[latin1]{inputenc}
\usepackage{geometry}
\usepackage{float}
\usepackage{mathrsfs}
\usepackage{amsmath}
\usepackage{amssymb}
\usepackage{graphicx}

\makeatletter

\floatstyle{ruled}
\newfloat{algorithm}{tbp}{loa}
\providecommand{\algorithmname}{Algorithm}
\floatname{algorithm}{\protect\algorithmname}

\usepackage{amsthm}

\usepackage{mathrsfs}

\usepackage{amsfonts}

\usepackage{epsfig}

\usepackage{bm}

\usepackage{mathrsfs}

\usepackage{enumerate}

\@ifundefined{definecolor}{\@ifundefined{definecolor}
 {\@ifundefined{definecolor}
 {\usepackage{color}}{}
}{}
}{}

\newtheorem{proposition}{Proposition}[section]

\newcounter{hypA}
\newenvironment{hypA}{\refstepcounter{hypA}\begin{itemize}
  \item[({\bf A\arabic{hypA}})]}{\end{itemize}}

\usepackage{babel}\date{}

\makeatother

\begin{document}

\title{Bayesian Parameter Inference for Partially Observed Stopped Processes}

\author{Ajay Jasra$^{1}$ and Nikolas Kantas$^{2}$}
\maketitle

\begin{center}



{\footnotesize $^{1}$Department of Statistics \& Applied Probability,
National University of Singapore, Singapore, 117546, Sg.}\\
{\footnotesize E-Mail:\,}\texttt{\emph{\footnotesize staja@nus.edu.sg}}\\
{\footnotesize $^{2}$Department of Electrical Engineering,
Imperial College London, London, SW7 2AZ, UK.}\\
{\footnotesize E-Mail:\,}\texttt{\emph{\footnotesize n.kantas@imperial.ac.uk}}
\end{center}

\begin{abstract}
In this article we consider Bayesian parameter inference associated
to partially-observed stochastic processes that start from a set $B_{0}$
and are stopped or killed at the first hitting time of a known set
$A$. Such processes occur naturally within the context of a wide
variety of applications. The associated posterior distributions are
highly complex and posterior parameter inference requires the use
of advanced Markov chain Monte Carlo (MCMC) techniques. Our approach
uses a recently introduced simulation methodology, particle Markov
chain Monte Carlo (PMCMC) \cite{pmcmc}, where sequential Monte Carlo
(SMC) \cite{doucet,liu} approximations are embedded within MCMC.
However, when the parameter of interest is fixed, standard SMC algorithms
are not always appropriate for many stopped processes. In \cite{chen,delmoral}
the authors introduce SMC approximations of multi-level Feynman-Kac
formulae, which can lead to more efficient algorithms. This is achieved
by devising a sequence of nested sets from $B_{0}$ to $A$ and then
perform the resampling step only when the samples of the process reach
intermediate level sets in the sequence. Naturally, the choice of
the intermediate level sets is critical to the performance of such
a scheme. In this paper, we demonstrate that multi-level SMC algorithms
can be used as a proposal in PMCMC. In addition, we propose a flexible
strategy that adapts the level sets for different parameter proposals.
Our methodology is illustrated on the coalescent model with migration. 
\end{abstract}

\section{Introduction\label{sec:intro}}

In this article we consider Markov processes that are stopped when
reaching the boundary of a given set $A$. These processes appear
in a wide range of applications, such as population genetics \cite{coal1,deiorio},
finance \cite{CasellaKilledDiff}, neuroscience \cite{bibbona}, physics
\cite{delmoral_garnier,johansen} and engineering \cite{blom,lezaudKrystulLGland}.
The vast majority of the papers in the literature deal with fully
observed stopped processes and assume the parameters of the model
are known. In this paper we address problems when this is not the
case. In particular, Bayesian inference for the model parameters is
considered, when the stopped process is observed indirectly via data.
We will propose a generic simulation method that can cope with many
types of partial observations. To the best of our knowledge, there
is no previous work in this direction. An exception is \cite{bibbona},
where maximum likelihood inference for the model parameters is investigated
for the fully observed case.

In the fully observed case, stopped processes have been studied predominantly
in the area of rare event simulation. In order to estimate the probability
of rare events related to stopped processes, one needs to efficiently
sample realisations of a process that starts in a set $B_{0}$ and
terminates in the given rare target set $A$ before returning to $B_{0}$
or getting trapped in some absorbing set. This is usually achieved
using Importance Sampling (IS) or multi-level splitting; see\cite{glasserman,sadowski_bucklew00}
and the references in those articles for an overview. Recently, sequential
Monte Carlo (SMC) methods based on both these techniques have been
used in \cite{blom,delmoral_garnier,johansen}. In \cite{cerou_nonasympt}
the authors also prove under mild conditions that SMC can achieve
same performance as popular competing methods based on traditional
splitting.

Sequential Monte Carlo methods can be described as a collection of
techniques used to approximate a sequence of distributions whose densities
are known point-wise up to a normalizing constant and are of increasing
dimension. SMC methods combine importance sampling and resampling
to approximate distributions. The idea is to introduce a sequence
of proposal densities and to sequentially simulate a collection of
$N\gg1$ samples, termed particles, in parallel from these proposals.
The success of SMC lies in incorporating a resampling operation to
control the variance of the importance weights, whose value would
otherwise increase exponentially as the target sequence progresses
e.g.~\cite{doucet,liu}.

Applying SMC in the context of fully observed stopped processes requires
using resampling while taking into account how close a sample is to
the target set. That is, it is possible that particles close to $A$
are likely to have very small weights, whereas particles closer to
the starting set $B_{0}$ can have very high weights. As a result,
the diversity of particles approximating longer paths before reaching
$A$ would be depleted by successive resampling steps. In population
genetics, for the coalescent model \cite{coal1}, this has been noted
as early as in the discussion of \cite{pg} by the authors of \cite{chen}.
Later, in \cite{chen} the authors used ideas from splitting and proposed
to perform the resampling step only when each sample of the process
reaches intermediate level sets, which define a sequence of nested
sets from $B_{0}$ to $A$. The same idea appeared in parallel in
\cite[Section 12.2]{delmoral}, where it was formally interpreted
as an interacting particle approximation of appropriate multi-level
Feynman-Kac formulae. Naturally, the choice of the intermediate level
sets is critical to the performance of such a scheme. That is, the
levels should be set in a {}``direction'' towards the set $A$ and
so that each level can be reached from the previous one with some
reasonable probability \cite{glasserman}. This is usually achieved
heuristically using trial simulation runs. Also more systematic techniques
exist: for cases where large deviations can be applied in \cite{dean}
the authors use optimal control and in \cite{cerou_old,cerou_static}
the level sets are computed adaptively on the fly using the simulated
paths of the process.

The contribution of this paper is to address the issue of inferring
the parameters of the law of the stopped Markov process, when the
process itself is a latent process and is only partially observed
via some given dataset. In the context of Bayesian inference one often
needs to sample from the posterior density of the model parameters,
which can be very complex. Employing standard Markov chain Monte Carlo
(MCMC) methods is not feasible, given the difficulty one faces to
sample trajectories of the stopped process. In addition, using SMC
for sequential parameter inference has been notoriously difficult;
see \cite{adronline1,statsci_review}. In particular, due to the successive
resampling steps the simulated past of the path of each particle will
be very similar to each other. This has been a long standing bottleneck
when static parameters $\theta$ are estimated online using SMC methods
by augmenting them with the latent state. These issues have motivated
the recently introduced particle Markov chain Monte Carlo (PMCMC)
\cite{pmcmc}. Essentially, the method constructs a Markov chain on
an extended state-space in the spirit of \cite{pseudo-marginal},
such that one may apply SMC updates for a latent process, i.e.~use
SMC approximations within MCMC. In the context of parameter inference
for stopped process this brings up the possibility of using the multi-level
SMC methodology as a proposal in MCMC. This idea to the best of our
knowledge has not appeared previously in the literature. The main
contributions made in this article are as follows: 
\begin{itemize}
\item When the sequence of level sets is fixed \emph{a priori}, the validity
of using multi-level SMC within PMCMC is verified. 
\item To enhance performance we propose a flexible scheme where the level
sets are adapted to the current parameter sample. The method is shown
to produce unbiased samples from the target posterior density. In
addition, we show both theoretically and via numerical examples how
the mixing of the PMCMC algorithm is improved when this adaptive strategy
is adopted. 
\end{itemize}
This article is structured as follows: in Section \ref{sec:Problem-Formulation}
we formulate the problem and present the coalescent as a motivating
example. In Section \ref{sec:stopping_time} we present multi-level
SMC for stopped processes. In Section \ref{sec:pmcmc} we detail a
PMCMC algorithm which uses multi-level SMC approximations within MCMC.
In addition, specific adaptive strategies for the levels are proposed,
which are motivated by some theoretical results that link the convergence
rate of the PMCMC algorithm to the properties of multi-level SMC approximations.
In Section \ref{sec:numerical} some numerical experiments for the
the coalescent are given. The paper is concluded in Section \ref{sec:summary}.
The proofs of our theoretical results can be found in the appendix.

\subsection{Notations\label{sub:Notations}}

The following notations will be used. A measurable space is written
as $(E,\mathcal{E})$, with the class of probability measures on $E$
written $\mathscr{P}(E)$. For $\mathbb{R}^{n}$, $n\in\mathbb{N}$
the Borel sets are $\mathscr{B}(\mathbb{R}^{n})$. For a probability
measure $\gamma\in\mathscr{P}(E)$ we will denote the density with
respect to an appropriate $\sigma$-finite measure $dx$ as $\overline{\gamma}(x)$.
The total variation distance between two probability measures $\gamma_{1},\gamma_{2}\in\mathscr{P}(E)$
is written as $\|\gamma_{1}-\gamma_{2}\|=\sup_{A\in\mathcal{E}}\left|\gamma_{1}(A)-\gamma_{2}(A)\right|$.
For a vector $(x_{i},\dots,x_{j})$, the compact notation $x_{i:j}$
is used; if $i>j$ $x_{i:j}$ is a null vector. For a vector $x_{1:j}$,
$|x_{1:j}|_{1}$ is the $\mathbb{L}_{1}-$norm. The convention $\prod_{\emptyset}=1$
is adopted. Also, $\min\{a,b\}$ is denoted as $a\wedge b$ and $\mathbb{I}_{A}(x)$
is the indicator of a set $A$. Let $E$ be a countable (possibly
infinite dimensional) state-space, then 
\[
\mathcal{S}(E)=\left\{ R=(r_{ij})_{i,j\in E}:\: r_{ij}\geq0,\sum_{l\in E}r_{il}=1\mbox{ and }\nu R=\nu\mbox{ for some }\nu=(\nu_{i})_{i\in E}\mbox{ with }\nu_{i}\geq0,\sum_{l\in E}\nu_{l}=1\right\} 
\]
 denotes the class of stochastic matrices which possess a stationary
distribution. In addition, we will denote as $e_{i}=(0,\dots,0,1,0,\dots,0)$
the $d$-dimensional vector whose $i^{th}$ element is $1$ and is
$0$ everywhere else. Finally, for the discrete collection of integers
we will use the notation $\mathbb{T}_{d}=\{1,\dots,d\}$.

\section{Problem Formulation\label{sec:Problem-Formulation}}

\subsection{Preliminaries}

Let $\theta$ be a parameter vector on $(\Theta,\mathscr{B}(\Theta))$,
$\Theta\subseteq\mathbb{R}^{d_{\theta}}$ with an associated prior
$\pi_{\theta}\in\mathscr{P}(\Theta)$. The stopped process $\{X_{t}\}_{t\geq0}$
is a $(E,\mathcal{E})-$valued discrete-time Markov process defined
on a probability space $(\Omega,\mathscr{F},\mathbb{P}_{\theta})$,
where $\mathbb{P}_{\theta}$ is a probability measure defined for
every $\theta\in\Theta$ such that for every $A\in\mathscr{F}$, $\mathbb{P}_{\theta}(A)$
is $\mathscr{B}(\Theta)-$measurable. For simplicity will we will
assume throughout the paper that the Markov process is homogeneous.
The state of the process $\{X_{t}\}_{t\geq0}$ begins its evolution
in a non empty set $B_{0}$ obeying an initial distribution $\nu_{\theta}:\: B_{0}\rightarrow\mathscr{P}(B_{0})$
and a Markov transition kernel $P_{\theta}:E\times\Theta\rightarrow\mathscr{P}(E)$.
The process is killed once it reaches a non-empty target set $A\subset B_{0}\in\mathscr{F}$
such that $\mathbb{P}_{\theta}(X_{0}\in B_{0}\setminus A)=1$. The
associated stopping time is defined as 
\[
\mathcal{T}=\inf\{t\geq0:X_{t}\in A\},
\]
 where it is assumed that $\mathbb{P}_{\theta}(\mathcal{T}<\infty)=1$
and $\mathcal{T}\in\mathcal{I}$, where $\mathcal{I}$ is a collection
of positive integer values related to possible stopping times.

In this paper we assume that we have no direct access to the state
of the process. Instead the evolution of the state of the process
generates a random observations' vector, which we will denote as $Y$.
The realisation of this observations' vector is denoted as $y$ and
assume that it takes value in some non empty set $F$. We will also
assume that there is no restriction on $A$ depending on the observed
data $y$, but to simplify exposition this will be omitted from the
notation.

In the context of Bayesian inference we are interested in the posterior
distribution: 
\begin{equation}
\pi(d\theta,dx_{0:\tau},\tau|y)\propto\gamma_{\theta}(dx_{0:\tau},y,\tau)\pi(d\theta),\label{eq:target}
\end{equation}
 where $\tau\in\mathcal{I}$ is the stopping time, $\pi_{\theta}$
is the prior distribution and $\gamma_{\theta}$ is the un-normalised
complete-data likelihood with the normalising constant of this quantity
being: 
\[
Z_{\theta}=\sum_{\tau\in\mathcal{I}}\int_{E^{\tau+1}}\gamma_{\theta}(dx_{0:\tau},y,\tau)dx_{0:\tau}.
\]
 The subscript on $\theta$ will be used throughout to explicitly
denote the conditional dependance on the parameter $\theta$. Given
the specific structure of the stopped processes one may write $\gamma_{\theta}$
as 
\begin{equation}
\gamma_{\theta}(dx_{0:\tau},y,\tau)=\xi_{\theta}(y|x_{0:\tau})\mathbb{I}_{(A^{c})^{\tau}\times A}(x_{0:\tau})\nu_{\theta}(dx_{0})\prod_{t=1}^{\tau}P_{\theta}(dx_{t}|x_{t-1}),\label{eq:unnormalised_likeli}
\end{equation}
 where $\xi_{\theta}:\Theta\times F\times\big(\bigcup_{\tau\in\mathcal{I}}\{\tau\}\times E^{\tau+1}\big)\rightarrow(0,1)$
is the likelihood of the data given the trajectory of the process.
Throughout, it will be assumed that for any $\theta\in\Theta$, $y\in F$,
$\gamma_{\theta}$ admits a density $\overline{\gamma}_{\theta}(x_{0:\tau},y,\tau)$
with respect to~a $\sigma-$finite measure $dx_{0:\tau}$ on $\overline{E}=\big(\bigcup_{\tau\in\mathcal{I}}\{\tau\}\times E^{\tau+1}\big)$
and the posterior and prior distributions $\pi,\: p$ admit densities
$\overline{\pi},\:\overline{p}$ respectively both defined with respect
to appropriate $\sigma-$finite dominating measures.

Note that (\ref{eq:target}) is expressed as an inference problem
for $(\theta,x_{0:\tau},\tau)$ and not only $\theta$. The overall
motivation originates from being able to design a MCMC that can sample
from $\pi$, which requires one to write the target (or an unbiased
estimate of it) up-to a normalizing constant \cite{pseudo-marginal}.
Still, our primary interest lies in Bayesian inference for the parameter
and this can be recovered by the marginals of $\pi$ with respect
to $\theta$. As it will become clear in Section \ref{sec:pmcmc}
the numerical overhead when augmenting the argument of the posterior
is necessary and we believe that the marginals with respect to $x_{0:\tau},\tau$
might also be useful by-products.

\subsection{Motivating example: the coalescent\label{sec:coal_model} }

\begin{figure}[h]
 \centering \setlength{\unitlength}{8cm} \begin{picture}(1,1) \put(.5,.75){\line(0,1){0.25}}
\put(0.25,.75){\line(1,0){.5}} \put(0.75,.75){\line(0,-1){0.75}}
\put(0.25,.75){\line(0,-1){0.25}} \put(0.125,.5){\line(1,0){0.25}}
\put(0.125,.5){\line(0,-1){0.5}} \put(0.375,.5){\line(0,-1){0.5}}
\multiput(-0.1,0.0)(0.1,0){12}{\line(1,0){0.02}} \multiput(-0.1,0.2)(0.1,0){12}{\line(1,0){0.02}}
\multiput(-0.1,0.5)(0.1,0){12}{\line(1,0){0.02}} \multiput(-0.1,0.5)(0.1,0){12}{\line(1,0){0.02}}
\multiput(-0.1,0.6)(0.1,0){12}{\line(1,0){0.02}} \multiput(-0.1,0.75)(0.1,0){12}{\line(1,0){0.02}}
\put(0.56,.90){$A$} \put(0.76,0.62){$A\rightarrow C$} \put(-0.05,0.21){$A\rightarrow C$}
\put(0.1,-0.05){$C$} \put(0.35,-0.05){$A$} \put(0.73,-0.05){$C$}

\put(1.1,0.0){$\tau=4$ stop $y=\{1,2\}$} \put(1.1,0.2){$t=3$
mutation} \put(1.1,0.5){$t=2$ split} \put(1.1,0.6){$t=1$ mutation}
\put(1.1,0.75){$t=0$ split}

\end{picture}

\caption{Coalescent model example: each of $\{A,C\}$ denote the possible genetic
type of observed chromosomes. In this example we have $d=2$ and $m=3$.
The tree propagates forward in time form the MRCA downwards by a sequence
of split and mutation moves. Arrows denote a mutation of one type
of a chromosome to another. The name of the process originates from
viewing the tree backwards in time (from bottom to top) where the
points where the graph join are coalescent events.}

\label{fig:coalgraph} 
\end{figure}
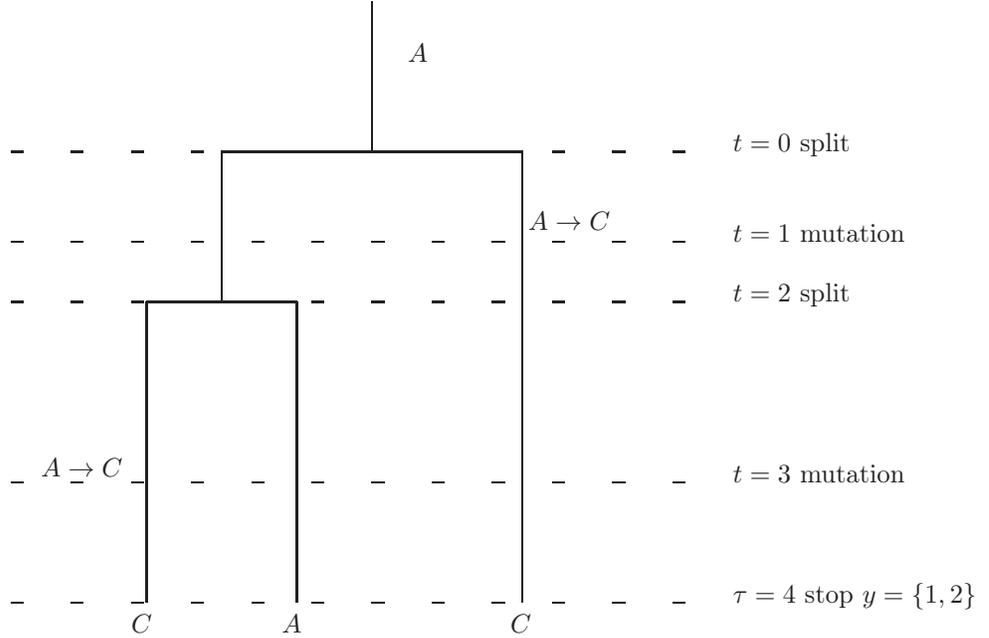

The framework presented so far is rather abstract, so we introduce
the coalescent model as a motivating example. In Figure \ref{fig:coalgraph}
we present a particular realisation of the coalescent for two genetic
types $\{A,C\}$. The process starts at epoch $t=0$ when the most
recent common ancestor (MRCA) splits into two versions of itself.
In this example $A$ is chosen to be the MCRA and the process continues
to evolve by split and mutation moves. At the stopping point (here
$t=4$) we observe some data $y$, which corresponds to the number
of genes for each genetic type.

In general we will assume there are $d$ different genetic types.
The latent state of the process $x_{t}^{i}$ is composed of the number
of genes of each type $i$ at epoch $t$ of the process and let also
$x_{t}=(x_{t}^{1},\dots,x_{t}^{d})$. The process begins by default
when the first split occurs, so the Markov chain $\left\{ X_{t}\right\} _{t\geq0}$
is initialised by the density 
\[
\overline{\nu}_{\theta}(x_{0})=\left\{ \begin{array}{ll}
\nu_{i} & \textrm{if}\quad x_{0}=2e_{i}\\
0 & \textrm{otherwise}.
\end{array}\right.
\]
 and is propagated using the following transition density: 
\[
\overline{P}_{\theta}(x_{t}|x_{t-1})=\left\{ \begin{array}{ll}
\frac{x_{t-1}^{i}}{|x_{t-1}|_{1}}\frac{\mu}{|x_{t-1}|_{1}-1+\mu}r_{il} & \textrm{if}\quad x_{t}=x_{t-1}-e_{i}+e_{l}\mbox{ (mutation)}\\
\frac{x_{t-1}^{i}}{|x_{t-1}|_{1}}\frac{|x_{t-1}|_{1}-1}{|x_{t-1}|_{1}-1+\mu} & \textrm{if}\quad x_{t}=x_{t-1}+e_{i}\mbox{ (split)}\\
0 & \textrm{otherwise},
\end{array}\right.
\]
 where $e_{i}$ is defined in Section \ref{sub:Notations}. Here the
first transition type corresponds to individuals changing type and
is called mutation, e.g. $A\rightarrow C$ at $t\in\{1,3\}$ in Figure
\ref{fig:coalgraph}. The second transition is called a split event,
e.g. $t\in\{0,2\}$ in the example of Figure \ref{fig:coalgraph}.
To avoid any confusion we stress that in Figure \ref{fig:coalgraph}
we present a particular realisation of the process that is composed
by a sequence of alternate split and mutations, but this is not the
only possible sequence. For example, the bottom of the tree could
have be obtained with $C$ being the possible MCRA and a sequence
of two consecutive splits and a mutation.

The process is stopped at epoch $\tau$ when the number of individuals
in the population reaches $m$. So for the state space we define:
\begin{eqnarray*}
\overline{E} & = & \bigcup_{t\in\mathcal{I}}\bigg(\{t\}\times E^{t+1}\bigg)\\
E & = & \{x:\: x\in(\mathbb{Z}^{+})^{d}\mbox{ and }2\leq|x|_{1}\leq m\}\\
\mathcal{I} & = & \{m,m+1,\dots\},
\end{eqnarray*}
 and for the initial and terminal sets we have: 
\begin{eqnarray*}
B_{0} & = & \{x:\: x\in\{0,2\}{}^{d}\mbox{ and }|x|_{1}=2\}\\
A & = & \{x:x\in(\mathbb{Z}^{+})^{d}\mbox{ and }|x|_{1}=m\}.
\end{eqnarray*}
 The data is generated by setting $y:=y^{1:d}=x_{\tau}\left(\in A\right)$,
which corresponds to the counts of genes that have been observed.
In the example of Figure \ref{fig:coalgraph} this corresponds to
$m=3$. Hence for the complete likelihood we have: 
\begin{equation}
\overline{\gamma}_{\theta}(x_{0:\tau},y,\tau)=\mathbb{I}_{A\cap\{x:x=y\}}(x_{\tau})\frac{\prod_{i=1}^{d}y^{i}!}{m!}\bigg[\overline{\nu}_{\theta}(x_{0})\prod_{t=1}^{\tau}\overline{P}_{\theta}(x_{t}|x_{t-1})\bigg]\label{eq:gamma_y}
\end{equation}
 As expected, the density is only non-zero if at time $\tau$ $x_{\tau}$
matches the data $y$ exactly.

Our objective is to infer the genetic parameters $\theta=(\mu,R)$,
where $\mu\in\mathbb{R}^{+}$ and $R\in\mathcal{S}(\mathbb{T}_{d})$
and hence the parameter space can be written as $\Theta=\mathbb{R}^{+}\times\mathcal{S}(\mathbb{T}_{d})$.
To facilitate Monte Carlo inference, one can reverse the time parameter
and simulate backward from the data. This is now detailed in the context
of importance sampling following the approach in \cite{griff}.

\subsubsection{Importance sampling for the coalescent model}

\label{sec:like_comp}

To sample realisations of the process for a given $\theta\in\Theta$,
importance sampling is adopted but with time reversed. First we introduce
a time reversed Markov kernel $M_{\theta}$ with density $\overline{M}_{\theta}(x_{t-1}|x_{t})$.
This is used as an importance sampling proposal where sampling is
performed backwards in time and the weighting forward in time. We
initialise using the data and simulate the coalescent tree backward
in time until two individuals remain of the same type. This procedure
ensures that the data is hit when the tree is considered forward in
time.

The process defined backward in time can be interpreted as a stopped
Markov process with the definitions of the initial and terminal sets
appropriately modified. For convenience we will consider the reverse
event sequence of the previous section, i.e we posed the problem backwards
in time with the reverse index being $j$. The proposal density for
the full path starting from the bottom of the tree and stopping at
its root can be written as 
\[
\overline{q}_{\theta}(x_{0:\tau})=\mathbb{I}_{B_{0}\cap\{x:x=y\}}(x_{0})\bigg\{\prod_{j=1}^{\tau}\overline{M}_{\theta}(x_{j}|x_{j-1})\bigg\}\mathbb{I}_{B_{0}}(x_{\tau}).
\]
 With reference to \eqref{eq:gamma_y} we have 
\[
\overline{\gamma}_{\theta}(x_{0:\tau},y,\tau)=\frac{m-1}{m-1+\mu}\frac{\prod_{i=1}^{d}y^{i}!}{m!}\overline{\nu}_{\theta}(x_{\tau})\bigg\{\prod_{j=1}^{\tau}\frac{\overline{P}_{\theta}(x_{j-1}|x_{j})}{\overline{M}_{\theta}(x_{j}|x_{j-1})}\bigg\}\overline{q}_{\theta}(x_{0:\tau}).
\]
 Then the marginal likelihood can be obtained 
\[
Z_{\theta}=\frac{m-1}{m-1+\mu}\frac{\prod_{i=1}^{d}y^{i}!}{m!}\sum_{\tau\in\mathcal{I}}\int_{E^{\tau+1}}\overline{\nu}_{\theta}(x_{\tau})\bigg\{\prod_{j=1}^{\tau}\frac{\overline{P}_{\theta}(x_{j-1}|x_{j})}{\overline{M}_{\theta}(x_{j}|x_{j-1})}\bigg\}\overline{q}_{\theta}(x_{0:\tau})dx_{0:\tau}.
\]
 In \cite{pg} the authors derive an optimal proposal $\overline{M}_{\theta}$
with respect to the variance of the marginal likelihood estimator.
For the sake of brevity we omit any further details. In the current
setup where there is only mutation and coalescences, the stopped-process
can be integrated out \cite{teh}, but this is not typically possible
in more complex scenarios. A more complicated problem, including migration,
is presented in Section \ref{sec:migration}. Finally, we remark that
the relevance of the marginal likelihood above will become clear later
in Section \ref{sec:pmcmc} as a crucial element in numerical algorithms
for inferring $\theta$.

\section{Multi-Level Sequential Monte Carlo Methods}

\label{sec:stopping_time}

In this section we shall briefly introduce generic SMC without extensive
details. We refer the reader for a more detailed description to \cite{delmoral,doucet}.
To ease exposition, when presenting generic SMC, we shall drop the
dependence upon parameter $\theta$.

SMC algorithms are designed to simulate from a sequence of probability
distributions $\pi_{1},\pi_{2},\dots,\pi_{p}$ defined on state space
of increasing dimension, namely $(G_{1},\mathscr{G}_{1}),(G_{1}\times G_{2},\mathscr{G}_{1}\otimes\mathscr{G}_{2}),\dots,(G_{1}\times\cdots\times G_{p},\mathscr{G}_{1}\otimes\cdots\otimes\mathscr{G}_{p})$.
Each distribution in the sequence is assumed to possess densities
with respect to a common dominating measure: 
\[
\overline{\pi}_{n}(u_{1:n})=\frac{\overline{\gamma}_{n}(u_{1:n})}{Z_{n}}
\]
 with each un-normalised density being $\overline{\gamma}_{n}:G_{1}\times\cdots\times G_{n}\rightarrow\mathbb{R}_{+}$
and the normalizing constant being $Z_{n}$. We will assume throughout
the article that there are natural choices for $\{\bar{\gamma}_{n}\}$
and that we can evaluate each $\overline{\gamma}_{n}$ point-wise.
In addition, we do not require knowledge of $Z_{n}.$

\subsection{Generic SMC algorithm}

\label{sec:smc_algo}

SMC algorithms approximate $\{\overline{\pi}_{n}\}_{n=1}^{p}$ recursively
by propagating a collection of properly weighted samples, called particles,
using a combination of importance sampling and resampling steps. For
the importance sampling part of the algorithm, at each step $n$ of
the algorithm, we will use general proposal kernels $M_{n}$ with
densities $\overline{M}_{n}$, which possess normalizing constants
that do not depend on the simulated paths. A typical SMC algorithm
is given in Algorithm \ref{alg:Generic-SMC-Algorithm} and we obtain
the following SMC approximations for $\pi_{n},$ 
\[
\pi_{n}^{N}(du_{1:n})=\sum_{j=1}^{N}\bar{W}_{n}^{(j)}\delta_{u_{1:n}^{(i)}}(du_{1:n})
\]
 and for the normalizing constant $Z_{n}$: 
\begin{equation}
\widehat{Z}_{n}=\prod_{k=1}^{n}\bigg\{\frac{1}{N}\sum_{j=1}^{N}W_{k}^{(j)}\bigg\}.\label{eq:normal_est_smc}
\end{equation}
 
\begin{algorithm}
Initialisation, $n=1$:

\qquad{}For $i=1,\dots,N$ 
\begin{enumerate}
\item Sample $u_{1}^{(i)}\sim\overline{M}_{1}$ . 
\item Compute weights 
\[
W_{1}^{(i)}=\frac{\overline{\gamma}_{1}(u_{1}^{(i)})}{\overline{M}_{1}(u_{1}^{(i)})},\:\bar{W}_{1}^{(i)}=\frac{W_{1}^{(i)}}{\sum_{j=1}^{N}W_{1}^{(j)}}.
\]

\end{enumerate}
For $n=2,\ldots,p$,

\qquad{}For $i=1,\dots,N$, 
\begin{enumerate}
\item Resampling: sample index $a_{n-1}^{i}\sim f(\cdot|\bar{W}_{n-1})$,
where $\bar{W}_{n-1}=(\bar{W}_{n-1}^{(1)},\dots,\bar{W}_{n-1}^{(N)})$. 
\item Sample $u_{n}^{(i)}\sim\overline{M}_{n}(\cdot|u_{1:n-1}^{(a_{n-1}^{i})})$
and set $u_{1:n}^{(i)}=(u_{1:n-1}^{(a_{n-1}^{i})},u_{n}^{(i)})$. 
\item Compute weights 
\[
W_{n}^{(i)}=w_{n}(u_{1:n}^{(i)})=\frac{\overline{\gamma}_{n}(u_{1:n}^{(i)})}{\overline{\gamma}_{n-1}(u_{1:n-1}^{(i)})\overline{M}_{n}(u_{n}^{(i)}|u_{1:n-1}^{(i)})},\:\bar{W}_{n}^{(i)}=\frac{W_{n}^{(i)}}{\sum_{j=1}^{N}W_{n}^{(j)}}.
\]

\end{enumerate}
\caption{\label{alg:Generic-SMC-Algorithm}Generic SMC Algorithm}
\end{algorithm}

In this paper we will use $f$ to be the multinomial distribution.
Then the resampled index of the ancestor of particle $i$ at time
$n$, namely $a_{n-1}^{i}\in\{1,\dots,N\}$, is also a random variable
with value chosen with probability $\bar{W}_{n-1}^{(a_{n-1}^{i})}$.
For each time $n$, we will denote the complete collection of ancestors
obtained from the resampling step as $\bar{\mathbf{a}}_{n}=(a_{n}^{1},\dots,a_{n}^{N})$
and the randomly simulated values of the state obtained during sampling
(step 2 for $n\geq2$) as $\bar{u}_{n}=(u_{n}^{(1)},\dots,u_{n}^{(N)})$.
We will also denote $\bar{\mathbf{a}}_{1:p},\bar{u}_{1:p}$ the concatenated
vector of all these variables obtained during the simulations from
time $n=1,\ldots,p$. Note $\bar{u}_{1:p}$ the is a vector containing
all $N\times p$ simulated states and should not be confused with
the particle sample of the path $(u_{1:p}^{(1)},\ldots,u_{1:p}^{(N)})$.

Furthermore, the joint density of all the sampled particles and the
resampled indices is 
\begin{equation}
\psi(\bar{u}_{1:p},\bar{\mathbf{a}}_{1:p-1})=\bigg(\prod_{i=1}^{N}\overline{M}_{1}(u_{1}^{(i)})\bigg)\prod_{n=2}^{p}\bigg(\prod_{i=1}^{N}\bar{W}_{n-1}^{(a_{n-1}^{i})}\overline{M}_{n}(u_{n}^{(i)}|u_{n-1}^{(a_{n-1}^{i})},\dots,u_{1}^{(a_{1}^{i})})\bigg),\label{eq:smc_algo}
\end{equation}
 The complete ancestral genealogy at each time can always traced back
by defining an ancestry sequence $b_{1:n}^{i}$ for every $i\in\mathbb{T}_{N}$
and $n\geq2$. In particular, we set the elements of $b_{1:n}^{i}$
using the backward recursion $b_{n}^{i}=a_{n}^{b_{n+1}^{i}}$ where
$b_{p}^{i}=i$. In this context one can view SMC approximations as
random probability measures induced by the imputed random genealogy
$\bar{\mathbf{a}}_{1:n}$ and all the possible simulated state sequences
that can be obtained using $\bar{u}_{1:n}$. This interpretation of
SMC approximations was introduced in \cite{pmcmc} and will be later
used together with $\psi(\bar{u}_{1:p},\bar{\mathbf{a}}_{1:p-1})$
for establishing the complex extended target distribution of PMCMC.

\subsection{Multi-Level SMC implementation}

\label{sec:stopping_time_smc}

For different classes of problems one can find a variety of enhanced
SMC algorithms; see e.g.~\cite{doucet}. In the context of stopped
processes, a multi-level SMC implementation was proposed in \cite{chen}
and the approach was illustrated for the coalescent model of Section
\ref{sec:coal_model}. We consider a modified approach along the lines
of Section 12.2 of \cite{delmoral} which seems better suited for
general stopped processes and can provably yield estimators of much
lower variance relative to vanilla SMC.

Introduce an arbitrary sequence of $\mathscr{F}-$nested sets 
\[
B_{0}\supset B_{1}\cdots\supset B_{p}=A,\:\quad p\geq2
\]
 with the corresponding stopping times denoted as 
\[
\mathcal{T}_{l}=\inf\{t\geq0:X_{t}\in B_{l}\},\:\quad1\leq l\leq p,
\]
 Note that the Markov property of $X_{t}$ implies~$0\leq\mathcal{T}_{1}\leq\mathcal{T}_{2}\leq\cdots\leq\mathcal{T}_{p}=\mathcal{T}$.

The implementation of multi-level SMC differs from the generic algorithm
of Section \ref{sec:smc_algo} in that between successive resampling
steps one proceeds by propagating in parallel trajectories of $X_{0:t}^{(j)}$
until the set $B_{n}$ is reached for each $j\in\mathbb{T}_{N}$.
For a given $j\in\mathbb{T}_{N}$ the path $X_{0:t}^{(j)}$ is {}``frozen''
once $X_{0:t}^{(j)}\in B_{n}$, until the remaining particles reach
$B_{n}$ and then a resampling step is performed. More formally denote
for $n=1$ 
\[
\mathcal{X}_{1}=\left(x_{0:\tau_{1}},\tau_{1}\right)\:\in\{x_{0:\tau_{1}},\tau_{1}:\: x_{0:\tau_{1}-1}\in B_{0}\setminus B_{1},\: x_{\tau_{1}}\in B_{1}\}
\]
 where $\tau_{1}$ is a realisation for the stopping time $T_{1}$
and similarly for $2\leq n\leq p$ we have 
\[
\mathcal{X}_{n}=\left(x_{\tau_{n-1}+1:\tau_{n}},\tau_{n}\right)\:\in\{x_{\tau_{n-1}+1:\tau_{n}},\tau_{n}:\: x_{\tau_{n-1}+1:\tau_{n}-1}\in B_{n-1}\setminus B_{n},\: x_{\tau_{n}}\in B_{n}\}.
\]

Multi-level SMC is a SMC algorithm which ultimately targets a sequence
of distributions $\{\pi_{n}\}$ each defined on a space 
\begin{equation}
\overline{E}_{n}=\bigcup_{\tau_{n}\in\mathcal{I}_{n}}\{\tau_{n}\}\times E^{\tau_{n}+1}\label{eq:sets}
\end{equation}
 where $n\in\mathbb{T}_{p}$, $p\geq2$ and $\mathcal{I}_{1},\ldots,\mathcal{I}_{p}$
are finite collections of positive integer values related to the stopping
times $\mathcal{T}_{1},\ldots,\mathcal{T}_{p}$ respectively. In the
spirit of generic SMC define intermediate target densities $\overline{\pi}_{n}$
w.r.t to an appropriate $\sigma$-finite dominating measure $d\mathcal{X}_{n}$.
We will assume there exists a natural sequence of densities $\{\overline{\pi}_{n}=\frac{\overline{\gamma}_{n}}{Z_{n}}\}_{1\leq n\leq p}$
obeying the restriction $\overline{\gamma}_{p}\equiv\overline{\gamma}_{\theta}$
so that the last target density $\overline{\gamma}_{p}$ coincides
with $\overline{\gamma}_{\theta}$ in (\ref{eq:unnormalised_likeli}).
Note that we define a sequence of $p$ target densities, but this
time the dimension of $\overline{\gamma}_{n}$ compared to $\overline{\gamma}_{n-1}$
grows with a random increment of $\tau_{n}-\tau_{n-1}$. In addition,
$\overline{\gamma}_{p}$ should clearly depend on the value of $\theta$,
but this suppressed in the notation. $ $The following proposition
is a direct consequence of the Markov property: \begin{proposition}
\label{prop:markov_level}Assume $\mathbb{P}_{\theta}(\mathcal{T}<\infty)=1$.
Then the stochastic sequence defined $(\mathcal{X}_{n})_{1\leq n\leq p}$
forms a Markov chain taking values in $\overline{E}_{n}$ defined
\eqref{eq:sets}. In addition, for any bounded measurable function
$h:\overline{E}_{n}\rightarrow\mathbb{R}$, then $\int_{\overline{E}_{n}}h(\mathcal{X}_{p})\gamma_{p}(d\mathcal{X}_{p})=\sum_{\tau\in\mathcal{I}}\int_{E^{\tau+1}}h(x_{0:\tau},\tau)\gamma_{\theta}(dx_{0:\tau},y,\tau)$.\end{proposition}
The proof can be found in \cite[Proposition 12.2.2, page 438]{delmoral},
\cite[Proposition 12.2.4, page 444]{delmoral} and the second part
is due to $\overline{\gamma}_{p}=\overline{\gamma}_{\theta}$.

We will present multi-level SMC based as a particular implementation
of the generic SMC algorithm. Firstly we replace $u_{n},u_{1:n}$
with $\mathcal{X}_{n},\mathcal{X}_{1:n}$ respectively. Contrary to
the presentation of Algorithm \ref{alg:Generic-SMC-Algorithm} for
multi-level SMC we will use a homogeneous Markov importance sampling
kernel $M_{\theta}(dx_{t}|x_{t-1})$, where $M_{\theta}:\Theta\times E\rightarrow\mathscr{P}(E)$,
$M_{\theta}(dx_{0}|x_{-1})\equiv M_{\theta}(dx_{0})$ by convention
and $\overline{M}_{\theta}$ is the corresponding density w.r.t. $dx$.
To compute the importance sampling weights of step 3 for $n\geq2$
in Algorithm \ref{alg:Generic-SMC-Algorithm} we use instead: 
\[
w_{n}(\mathcal{X}_{1},\dots,\mathcal{X}_{n})=\frac{\overline{\gamma}_{n}(\mathcal{X}_{1},\dots,\mathcal{X}_{n})}{\overline{\gamma}_{n-1}(\mathcal{X}_{1},\dots,\mathcal{X}_{n-1})\prod_{l=\tau_{n-1}+1}^{\tau_{n}}\overline{M}_{\theta}(x_{l}|x_{l-1})}.
\]
 and for step 2 at $n=1$: 
\[
w_{1}(\mathcal{X}_{1})=\frac{\overline{\gamma}_{1}(\mathcal{X}_{1})}{\prod_{l=0}^{\tau_{1}}\overline{M}_{\theta}(x_{l}|x_{l-1})}.
\]
 To simplify notation from herein we write 
\[
\mathcal{M}_{1}(\mathcal{X}_{1})=\prod_{l=0}^{\tau_{1}}\overline{M}_{\theta}(x_{l}|x_{l-1})
\]
 and given $p$, for any $2\leq n\leq p$ we have 
\[
\mathcal{M}_{n}(\mathcal{X}_{n}|\mathcal{X}_{n-1})=\prod_{l=\tau_{n-1}+1}^{\tau_{n}}\overline{M}_{\theta}(x_{l}|x_{l-1}),
\]
where again we have suppressed the $\theta$-dependance of $\mathcal{M}_{n}$
in the notation. We present the multi-level SMC algorithm in Algorithm
\ref{alg:Multilevel-SMC-Algorithm}. Note here we include a procedure
whereby at each stage $n$, particles that do not reach $B_{n}$ before
time $t_{n}$ are rejected by assigning them a zero weight, whereas
before it was hinted that resampling is performed when all particles
reach $B_{n}$. Similar to \eqref{eq:smc_algo}, it is clear that
the joint probability density of all the random variables used to
implement a multi-level SMC algorithm with multinomial resampling
is given by: 
\begin{equation}
\psi_{\theta}(\bar{\mathcal{X}}_{1:p},\bar{\mathbf{a}}_{1:p-1})=\bigg(\prod_{i=1}^{N}\mathcal{M}_{1}(\mathcal{X}_{1}^{(i)})\bigg)\prod_{n=2}^{p}\bigg(\prod_{i=1}^{N}\bar{W}_{n-1}^{(a_{n-1}^{i})}\mathcal{M}_{n}(\mathcal{X}_{n}^{(i)}|\mathcal{X}_{n-1}^{(a_{n-1}^{i})})\bigg),\label{eq:stopping_density}
\end{equation}
 where $\bar{\mathcal{X}}_{1:p}$ is defined similarly to $\bar{u}_{1:p}$.
Finally, recall by construction $Z_{p}=Z_{\theta}$ so the approximation
of the normalizing constant of $\gamma_{\theta}$ for a fixed $\theta$
is 
\begin{equation}
\widehat{Z}_{\theta}=\widehat{Z}_{p}=\prod_{n=1}^{p}\bigg\{\frac{1}{N}\sum_{j=1}^{N}w_{n}^{(j)}\left(\mathcal{X}_{1:n}^{(j)}\right)\bigg\}.\label{eq:norm_est_stop}
\end{equation}
 
\begin{algorithm}
Initialisation, $n=1$:

\qquad{}For $i=1,\dots,N$ 
\begin{enumerate}
\item For $t=1,\ldots,t_{1}$: 

\begin{enumerate}
\item Sample $x_{t}^{(i)}\sim M_{\theta}(\cdot|x_{t-1}^{(i)})$. 
\item If $x_{t}^{(i)}\in B_{1}$ set $\tau_{1}^{(i)}=t$ , $\mathcal{X}_{1}^{(i)}=\left(x_{0:\tau_{1}^{(i)}}^{(i)},\tau_{1}^{(i)}\right)$
and go to step 2.
\end{enumerate}
\item Compute weights 
\[
W_{1}^{(i)}=\frac{\overline{\gamma}_{1}(\mathcal{X}_{1}^{(i)})\mathbb{I}_{\tau_{1}^{(i)}\leq t_{1}}}{\mathcal{M}_{1}(\mathcal{X}_{1}^{(i)})},\:\bar{W}_{1}^{(i)}=\frac{W_{1}^{(i)}}{\sum_{j=1}^{N}W_{1}^{(j)}}.
\]

\end{enumerate}
For $n=2,\ldots,p$,

\qquad{}For $i=1,\dots,N$, 
\begin{enumerate}
\item Resampling: sample index $a_{n-1}^{i}\sim f(\cdot|\bar{W}_{n-1})$,
where $\bar{W}_{n-1}=(\bar{W}_{n-1}^{(1)},\dots,\bar{W}_{n-1}^{(N)})$. 
\item For $t=\tau_{n-1}^{(i)}+1,\ldots,t_{n}$: 

\begin{enumerate}
\item Sample $x_{t}^{(i)}\sim M_{\theta}(\cdot|x_{t-1}^{(i)})$. 
\item If $x_{t}^{(i)}\in B_{n}$ set $\tau_{n}^{(i)}=t$ , $\mathcal{X}_{n}^{(i)}=\left(x_{\tau_{n-1}^{(i)}+1:\tau_{n}^{(i)}}^{(i)},\tau_{n}^{(i)}\right)$
and go to step 3.
\end{enumerate}
\item Set $\mathcal{X}_{1:n}^{(i)}=(\mathcal{X}_{1:n-1}^{(a_{n-1}^{i})},\mathcal{X}_{n}^{(i)})$. 
\item Compute weights 
\[
W_{n}^{(i)}=w_{n}(\mathcal{X}_{1:n}^{(i)})=\frac{\overline{\gamma}_{n}(\mathcal{X}_{1:n}^{(i)})\mathbb{I}_{\tau_{n}^{(i)}\leq t_{n}}}{\overline{\gamma}_{n-1}(\mathcal{X}_{1:n-1}^{(i)})\mathcal{M}_{n}(\mathcal{X}_{n}^{(i)}|\mathcal{X}_{1:n-1}^{(i)})},\:\bar{W}_{n}^{(i)}=\frac{W_{n}^{(i)}}{\sum_{j=1}^{N}W_{n}^{(j)}}.
\]

\end{enumerate}
\caption{\label{alg:Multilevel-SMC-Algorithm}Multi-level SMC Algorithm}
\end{algorithm}

\subsubsection{Setting the levels }

\label{sec:problems_stopping_smc}

We will begin by showing how the levels can be set for the coalescent
example of Section \ref{sec:coal_model}. We will proceed in the spirit
of Section \ref{sec:like_comp} and consider the backward process
so that the {}``time'' indexing is set to start from the bottom
of the tree towards the root. We introduce a a collection of integers
$m>l_{1}>l_{2}>\cdots>l_{p}=2$ and define 
\begin{align*}
B_{0} & =\{x\in(\mathbb{Z}^{+}\cup\{0\})^{d}:\: x=y\},\: n=0,\\
B_{n} & =\{x\in(\mathbb{Z}^{+}\cup\{0\})^{d}:\:|x|_{1}=l_{n}\},\:1\leq n\leq p.
\end{align*}
 Clearly we have $B_{n-1}\supset B_{n}$, $1\leq n\leq p$ and $B_{p}=A$.
One can also write the sequence of target densities for the multi-level
setting as: 
\begin{align*}
\overline{\gamma}_{1}(x_{0:\tau_{1}},\tau_{1}) & =\frac{m-1}{m-1+\mu}\frac{\prod_{i=1}^{d}(y^{i})!}{m!}\mathbb{I}_{\{y\}}(x_{0})\prod_{l=1}^{\tau_{1}}\overline{P}_{\theta}(x_{l-1}|x_{l})\mathbb{I}_{\{x_{t_{n}}\in B_{n}\}}(x_{t_{n}}),\\
\overline{\gamma}_{n}(x_{0:\tau_{n}},\tau_{n}) & =\overline{\gamma}_{n-1}(x_{0:\tau_{n-1}},\tau_{n-1})\prod_{l=\tau_{n-1}+1}^{\tau_{n}}\overline{P}_{\theta}(x_{l-1}|x_{l})\mathbb{I}_{\{x_{t_{n}}\in B_{n}\}}(x_{t_{n}}),\: n=1,\dots,p.
\end{align*}

The major design problem that remains in general is that given \emph{any}
candidates for $\{\overline{M}_{n,\theta}\}$, how to set the spacing
(in some sense) of the $\{B_{n}\}$ and how many levels are needed
so that good SMC algorithms can be constructed. That is, if the $\{B_{n}\}$
are far apart, then one can expect that weights will degenerate very
quickly and if the $\{B_{n}\}$ are too close that the algorithm will
resample too often and hence lead to poor estimates. For instance,
in the context of the coalescent example of Section \ref{sec:coal_model},
if one uses the above construction for $\{B_{n}\}$ the importance
weight at the $n$-th resampling time is 
\[
w_{n}(x_{0:\tau_{n}})=\prod_{l=\tau_{n-1}+1}^{\tau_{n}}\frac{\overline{P}_{\theta}(x_{l-1}|x_{l})}{\overline{M}_{\theta,n}(x_{l}|x_{l-1})}\mathbb{I}_{\{x_{\tau_{n}}\in B_{n}\}}(x_{\tau_{n}}),
\]
 Now, in general for any $\{l_{n}\}_{n=1}^{p}$ and $p$ it is hard
to know beforehand how much better (or not) the resulting multi-level
algorithm will perform relative to a vanilla SMC algorithm. Whilst
\cite{chen} show empirically that in most cases one should expect
a considerable improvement, there $\theta$ is considered to be fixed.
In this case one could design the levels sensibly using offline heuristics
or more advanced systematic methods using optimal control \cite{dean}
or adaptive simulation \cite{cerou_old,cerou_static}, e.g. by setting
the next level using the median of a pre-specified rank of the particle
sample. What we aim to establish in the next section is that when
$\theta$ is varies as in the context of MCMC algorithms, one can
both construct PMCMC algorithms based on multi-level SMC and more
importantly easily design for each $\theta$ different sequences for
$\{B_{n}\}$ based on similar ideas.

\section{Multi-Level Particle Markov Chain Monte Carlo\label{sec:pmcmc}}

Particle Markov Chain Monte Carlo (PMCMC) methods are MCMC algorithms,
which use all the random variables generated by SMC approximations
as proposals. As in standard MCMC the idea is to run an ergodic Markov
chain to obtain samples from the distribution of interest. The difference
lies that in order use the simulated variables from SMC, one defines
a complex invariant distribution for the MCMC on an extended state
space. This extended target is such that a marginal of this invariant
distribution is the one of interest.

This section aims on providing insight to the following questions: 
\begin{enumerate}
\item {Is it valid in general to use multi-level SMC within PMCMC?} 
\item {Given that it is, how can we use the levels to improve the mixing
of PMCMC?} 
\end{enumerate}
The answer to the first question seems rather obvious, so we will
provide some standard but rather strong conditions for which multi-level
PMCMC is valid. For the second question we will propose an extension
to PMMH that adapts the level sets used to $\theta$ at every iteration
of PMCMC. \cite{pmcmc} introduces three different and generic PMCMC
algorithms: particle independent Metropolis Hastings algorithm (PIMH),
particle marginal Metropolis Hastings (PMMH) and particle Gibbs samplers.
In the remainder of the paper we will only focus on the first two
of these.

\subsection{Particle independent Metropolis Hastings (PIMH) \label{sec:pmcmc_meth}}

\begin{algorithm}
\begin{enumerate}
\item Sample $\bar{\mathcal{X}}_{1:p},\bar{\mathbf{a}}_{1:p-1}$ from \eqref{eq:stopping_density}
using the multi-level implementation of Algorithm \ref{alg:Generic-SMC-Algorithm}
detailed in Section \eqref{sec:stopping_time_smc} and compute $\widehat{Z}_{p}$.
Sample $k\sim f(\cdot|\bar{W}_{p})$ . 
\item Set $\xi(0)=\left(k(0),\bar{\mathcal{X}}_{1:p}(0),\bar{\mathbf{a}}_{1:p-1}(0)\right)=\left(k,\bar{\mathcal{X}}_{1:p},\bar{\mathbf{a}}_{1:p-1}\right)$
and $\widehat{Z}_{p}(0)=\widehat{Z}_{p}.$ 
\item For $i=1,\dots,K$:

\begin{enumerate}
\item Propose a new $\bar{\mathcal{X}'}_{1:p},\bar{\mathbf{a}}'_{1:p}$
and $k'$ as in step 1 and compute $\widehat{Z}'_{p},$ 
\item Accept this as the new state of the chain with probability $1\wedge\frac{\widehat{Z_{p}}'}{\widehat{Z}_{p}(i-1)}.$
If we accept, set $\xi(i)=\left(k(i),\bar{\mathcal{X}}_{1:p}(i),\bar{\mathbf{a}}_{1:p-1}(i)\right)=\left(k',\bar{\mathcal{X}}'_{1:p},\bar{\mathbf{a}}'_{1:p-1}\right)$
and $\widehat{Z}_{p}(i)=\widehat{Z}'_{p}$. Otherwise reject, $\xi(i)=\xi(i-1)$
and $\widehat{Z}_{p}(i)=\widehat{Z}_{p}(i-1)$. 
\end{enumerate}
\end{enumerate}
\caption{Particle independent Metropolis-algorithm (PIMH)\label{alg:PIMH}}
\end{algorithm}

We will begin by presenting the simplest generic algorithm found in
\cite{pmcmc}, namely the particle independent Metropolis Hastings
algorithm (PIMH). In this case $\theta$ and $p$ are fixed and PIMH
is designed to sample from the pre-specified target distribution $\pi_{p}$
also considered in Section \ref{sec:smc_algo}. Although PIMH is not
useful for parameter inference it is included for pedagogic purposes.
One must bear in mind that PIMH is the most basic of all PMCMC algorithms.
As such it is easier to analyse but still can provide useful intuition
that can be used later in the context of PMMH and varying $\theta$.

PIMH is presented in Algorithm \ref{alg:PIMH}. It can be shown, using
similar arguments to \cite{pmcmc}, that the invariant density of
the Markov kernel above is exactly (see the proof of Proposition \ref{prop:stop_within_mcmc})
\[
\overline{\pi}_{p}^{N}(k,\bar{\mathcal{X}}_{1:p},\bar{\mathbf{a}}_{1:p-1})=\frac{1}{N^{p}}\frac{\gamma_{p}(\mathcal{X}_{1:p}^{(k)})}{Z_{p}}\frac{\psi_{\theta}(\bar{\mathcal{X}}_{1:p},\bar{\mathbf{a}}_{1:p-1})}{\mathcal{M}_{1}(\mathcal{X}_{1}^{(b_{1}^{k})})\prod_{n=2}^{p}\big\{\bar{W}_{n-1}^{(b_{n-1}^{k})}\mathcal{M}_{n}(\mathcal{X}_{n}^{(b_{n}^{k})}|\mathcal{X}_{n-1}^{(b_{n-1}^{k})})\big\}}
\]
 where $\psi$ is as in \eqref{eq:stopping_density} and as before
we have $b_{p}^{k}=k$ and $b_{n}^{k}=a_{n}^{b_{n+1}^{k}}$ for every
$k,n$ . Note that $\overline{\pi}_{p}^{N}$ admits the target density
of interest, $\overline{\pi}_{p}$ as the marginal, when $k$ and
$\bar{\mathbf{a}}_{1:p-1}$ are integrated out.

We commence by briefly investigating some convergence properties of
PIMH with multi-level SMC. Even though the scope of PIMH is not parameter
inference, one can use insight on what properties are desired by multi-level
SMC for PIMH when designing other PMCMC algorithms used for parameter
inference. We begin with posing the following mixing and regularity
assumption:

\begin{hypA}\label{assump:a1} For every $\theta\in\Theta$ and $p\in\mathcal{I}$
there exist a $\varphi\in(0,1)$ such that for every $(x,x')\in E\times E$:
\[
\varphi\leq\overline{M}_{\theta}(x'|x)\leq\varphi^{-1}
\]
 There exist a $\rho\in(0,1)$ such that for $1\leq n\leq p$ and
every $\mathcal{X}_{1:n}\in\bar{E}_{n}$: 
\[
\rho^{\tau_{n}}\leq\overline{\gamma}_{n}(\mathcal{X}_{1:n})\leq\rho^{-\tau_{n}}.
\]
 The stopping times are finite, that is for $1\leq n\leq p$ there
exist a $\bar{\tau}_{n}<\infty$ such that 
\[
\tau_{n}\leq\bar{\tau}_{n}.
\]
 \end{hypA}

Assumption (A\ref{assump:a1}) is rather strong, but are often used
in the analysis of these kind of algorithms \cite{pmcmc,delmoral}
because they simplify the proofs to a large extent. Recall that $\theta\in\Theta$
and $p\in\mathcal{I}$ are fixed. We proceed by stating the following
proposition:

\begin{proposition}\label{prop:conv_rate} Assume (A\ref{assump:a1}).
Then for $N\geq1$ Algorithm \ref{alg:PIMH} generates a sequence
$\left(\mathcal{X}_{1:p}(i)\right)_{i\geq0}$ that for any $i\geq1$,
$\xi(0)\in\mathbb{T}_{N}^{(p-1)N+1}\times\overline{E}$, $\theta\in\Theta$
satisfies: 
\[
\|\mathcal{L}aw(\mathcal{X}_{1:p}(i)\in\cdot|\xi(0))-\pi_{p}(\cdot)\|\leq\bigg(1-Z_{p}\left(\left(\rho\varphi\right){}^{2\sum_{j=1}^{p}\bar{\tau}_{j}}\right)\bigg)^{i}.
\]
 \end{proposition}The proof can be found in the appendix. The following
remarks are generalised and do not always hold, but provide some intuition
for the ideas that follow. The result shows intrinsically that as
the supremum of the sum of the stopping times with respect to $\{B_{n}\}_{n=1}^{p}$
gets smaller, so does the convergence rate increase. This can be also
linked to the variance of the estimator of $\widehat{Z}_{p}$, which
is well known to increase linearly with $p$ \cite[Theorem 12.2.2, pages 451-453]{delmoral}.
Shorter stopping times will typically yield lower variance and hence
better MCMC convergence properties. On the other hand often $\gamma_{p}$
will be larger for longer $p$ and longer stopping times (Proposition
\ref{prop:conv_rate} is derived for a fixed $p$). In addition, sampling
a stopped process is easier using a higher number of levels. In summary,
the tradeoff is that although it is more convenient to use more auxiliary
variables for simulating the process, these will slow down the mixing
of PMCMC. In practice one balances this by trying to use a moderate
number of levels for which most the particles to reach $A$. This
tradeoff serves as a motivation for developing flexible schemes to
vary $p,\{B_{n}\}_{n=1}^{p}$ with $\theta$ in the PMCMC algorithm
presented later in Section \ref{sec:adaptive_strat}.

\subsection{Particle marginal Metropolis Hastings (PMMH)}

In the remainder of this section we will focus on using a multi-level
SMC implementation within a PMMH algorithm. Given the commentary in
Section \ref{sec:stopping_time_smc} and our interest in drawing inference
on $\theta\in\Theta$, it seems that using multi-level SMC within
PMCMC should be highly beneficial. Recall \eqref{eq:target} can be
expressed in terms of densities as:

\begin{equation}
\overline{\pi}(\theta,\mathcal{X}_{1:p})\propto\overline{\gamma}_{p}(\mathcal{X}_{1:p})\overline{p}(\theta)\label{eq:target_den}
\end{equation}
 and let the marginal density given by 
\[
\bar{\pi}(\theta)=\sum_{\tau\in\mathcal{I}}\int_{E^{\tau+1}}\bar{\pi}(\theta,x_{0:\tau},\tau|y)dx_{0:\tau}.
\]
 For the time being we will consider the case when $p$ is fixed.
In the context of our stopped Markov process, we propose a PMMH algorithm
targeting $\overline{\pi}(\theta,\mathcal{X}_{1:p})$ in Algorithm
\ref{fig:stop_within_mcmc}.

\begin{algorithm}
\begin{enumerate}
\item Sample $\theta(0)\sim p(\cdot)$. Given $\theta(0)$ sample $\bar{\mathcal{X}}_{1:p}(0),\bar{\mathbf{a}}_{1:p-1}(0)$
using multi-level SMC and compute $\widehat{Z}_{\theta(0)}$. Sample
$k\sim f(\cdot|\bar{W}_{p})$ . 
\item Set $\xi(0)=\left(\theta(0),k(0),\bar{\mathcal{X}}_{1:p}(0),\bar{\mathbf{a}}_{1:p-1}(0)\right)$
and $\widehat{Z}_{\theta}(0)=\widehat{Z}_{\theta(0)}$ . 
\item For $i=1,\dots,K$:

\begin{enumerate}
\item Sample $\theta^{\prime}\sim q(\cdot|\theta(i-1))$; given $\theta^{\prime}$
propose a new $\bar{\mathcal{X}'}_{1:p},\bar{\mathbf{a}}'_{1:p-1}$
and $k'$ as in step 1 and compute $\widehat{Z}'_{\theta'}$. 
\item Accept this as the new state of the chain with probability 
\[
1\wedge\frac{\widehat{Z}'_{\theta^{\prime}}\bar{p}(\theta')}{\widehat{Z}_{\theta}(i-1)\bar{p}(\theta)}\times\frac{\bar{q}(\theta(i-1)|\theta^{\prime})}{\bar{q}(\theta^{\prime}|\theta(i-1))}.
\]
 If we accept, set $\xi(i)=\left(\theta(i),k(i),\bar{\mathcal{X}}_{1:p}(i),\bar{\mathbf{a}}_{1:p-1}(i)\right)=\left(\theta',k',\bar{\mathcal{X}}'_{1:p},\bar{\mathbf{a}}'_{1:p-1}\right)$
and $\widehat{Z}_{\theta}(i)=\widehat{Z}'_{\theta'}$. Otherwise reject,
$\xi(i)=\xi(i-1)$ and $\widehat{Z}_{\theta}(i)=\widehat{Z}_{\theta}(i-1)$. 
\end{enumerate}
\end{enumerate}
\caption{Particle marginal Metropolis Hastings using multi-Level SMC.\label{fig:stop_within_mcmc} }
\end{algorithm}

We will establish the invariant density and convergence of this algorithm,
under the following assumption:

\begin{hypA}\label{assump:a5-6} For any $\theta\in\Theta$ and $p\in\mathcal{I}$
we define the following sets for $n=1,\ldots,p$: $S_{n}^{\theta}=\{\mathcal{X}_{1:n}\in\overline{E}_{n}:\gamma_{n}(\mathcal{X}_{1:n})>0\}$
and $Q_{n}^{\theta}=\{\mathcal{X}_{1:n}\in\overline{E}_{n}:\gamma_{n-1}(\mathcal{X}_{1:n-1})\mathcal{M}_{\theta,n}(\mathcal{X}_{n}|\mathcal{X}_{n-1})>0\}$.
For any $\theta\in\Theta$ we have that $S_{n}^{\theta}\subseteq Q_{n}^{\theta}$.
In addition the ideal Metropolis Hastings targeting $\overline{\pi}(\theta)$
using proposal density $q(\theta'|\theta)$ is irreducible and aperiodic.
\end{hypA} This assumption contains Assumptions 5 and 6 of \cite{pmcmc}
modified to our problem with a simple change of notations. We proceed
with the following result:

\begin{proposition}\label{prop:stop_within_mcmc} Assume (A\ref{assump:a5-6});
then for any $N\geq1$: 
\begin{enumerate}
\item {The invariant density of the procedure described in Algorithm \ref{fig:stop_within_mcmc},
is on the space $\Theta\times\mathbb{T}_{N}^{(p-1)N+1}\times\overline{E}_{n}$
and has the representation 
\begin{equation}
\overline{\pi}_{p}^{N}(\theta,k,\bar{\mathcal{X}}_{1:p},\bar{\mathbf{a}}_{1:p-1})=\frac{\overline{\pi}(\theta,\mathcal{X}_{1:p}^{(k)})}{N^{p}}\frac{\psi_{\theta}(\bar{\mathcal{X}}_{1:p},\bar{\mathbf{a}}_{1:p-1})}{\mathcal{M}_{1}(\mathcal{X}_{1}^{(b_{1}^{k})})\prod_{n=2}^{p}\big\{\bar{W}_{n-1}^{(b_{n-1}^{k})}\mathcal{M}_{n}(\mathcal{X}_{n}^{(b_{n}^{k})}|\mathcal{X}_{n-1}^{(b_{n-1}^{k})})\big\}}\label{eq:stop_target}
\end{equation}
 where $\overline{\pi}$ is as in \eqref{eq:target_den} and $\psi_{\theta}$
as in \eqref{eq:stopping_density}. In addition, \eqref{eq:stop_target}
admits $\overline{\pi}(\theta)$ as a marginal. } 
\item {Algorithm \ref{fig:stop_within_mcmc} generates a sequence $\left(\theta(i),\mathcal{X}_{1:p}(i)\right){}_{i\geq0}$
such that 
\[
\lim_{i\rightarrow\infty}\|\mathcal{L}aw(\theta(i),\mathcal{X}_{1:p}(i)\in\cdot)-\pi(\cdot)\|=0
\]
 where $\pi$ is as in \eqref{eq:target}. } 
\end{enumerate}
\end{proposition}

The proof of the result is in the Appendix. The result is based on
Theorem 4 of \cite{pmcmc}. Note that Algorithm \ref{fig:stop_within_mcmc}
presented in a generic form of a {}``vanilla'' PMMH algorithm, so
it can be enhanced using various strategies. For example, it is possible
to add block updating of the latent variables or backward simulation
in the context of a particle Gibbs version \cite{whiteley}. In the
next section, we propose a flexible scheme that allows to set a different
number of levels after a new $\theta'$ $ $is proposed.

\subsection{Adapting the level sets}

\label{sec:adaptive_strat}

The remaining design issue for PMMH is how to tune multi-level SMC
by choosing $p$ and $\{B_{n}\}_{n=1}^{p}$. Whilst, for a fixed $\theta\in\Theta$,
one could solve the problem with preliminary runs, when $\theta$
varies this is not an option. In general the value of $\theta$ should
dictate how small or large $p$ should be to facilitate an efficient
SMC algorithm. Hence, to obtain a more accurate estimate of the marginal
likelihood and thus an efficient MCMC algorithm, we need to consider
adaptive strategies to propose randomly a different number of levels
$p$ and levels' sequence $\{B_{n}\}_{n=1}^{p}$ for each $\theta(i)$
sampled at every PMMH iteration $i$. To ease exposition we will assume
that $p,\{B_{n}\}_{n=1}^{p}$ can be expressed as functions of an
arbitrary auxiliary parameter $v$.

Given $\theta(i)$ is a random variable, the main questions we wish
to address is how to perform such an adaptive strategy consistently.
An important point, is the fact that since we are interested in parameter
inference, it is required that the marginal of the PMMH invariant
density is $\overline{\pi}(\theta)$. This can be ensured (see Proposition
\ref{prop:adap_stop}) by introducing at each PMMH iteration, the
parameters that form the level sets $v(i)$ as an auxiliary process,
which given $\theta(i)$ is conditionally independent of $k(i),\bar{\mathcal{X}}_{1:p}(i),\bar{\mathbf{a}}_{1:p-1}(i)$.
This way we define an extended target for the MCMC algorithm, which
includes $p$ and $\{B_{n}\}_{n=1}^{p}$ in the target variables.
It should be noted that this scheme is explicitly different from Proposition
1 of \cite{roberts1}, where the MCMC transition kernel at iteration
$i$ is dependent upon an auxiliary process. Here one just augments
the target space with more auxiliary variables.

\begin{algorithm}
\begin{enumerate}
\item Sample $\theta(0)\sim p(\cdot)$. Given $\theta(0)$: sample $v(0)$
from $\Lambda_{\theta(0)}$, then $\bar{\mathcal{X}}_{1:p(v(0))}(0),\bar{\mathbf{a}}_{1:p(v(0))-1}(0)$
using multi-level SMC and compute $\widehat{Z}_{\theta(0)}$. Sample
$k\sim f(\cdot|\bar{W}_{p})$ . 
\item Set $\xi(0)=\left(\theta(0),v(0),k(0),\bar{\mathcal{X}}_{1:p}(0),\bar{\mathbf{a}}_{1:p-1}(0)\right)$
and $\widehat{Z}_{\theta}(0)=\widehat{Z}_{\theta(0)}$ . 
\item For $i=1,\dots,K$:

\begin{enumerate}
\item Sample $\theta^{\prime}\sim q(\cdot|\theta(i-1))$; sample $v'$ from
$\Lambda_{\theta'}$ and $\bar{\mathcal{X}'}_{1:p(v')},\bar{\mathbf{a}}'_{1:p(v')-1},$
$k'$ as in step 1 and compute $\widehat{Z}'_{\theta'}$. 
\item Accept this as the new state of the chain with probability 
\[
1\wedge\frac{\widehat{Z}'_{\theta^{\prime}}\bar{p}(\theta')}{\widehat{Z}_{\theta}(i-1)\bar{p}(\theta)}\times\frac{\bar{q}(\theta(i-1)|\theta^{\prime})}{\bar{q}(\theta^{\prime}|\theta(i-1))}.
\]
 If we accept, set $\xi(i)=\left(\theta(i),v(i),k(i),\bar{\mathcal{X}}_{1:p(v(i))}(i),\bar{\mathbf{a}}_{1:p(v(i))-1}(i)\right)=\left(\theta',v',k',\bar{\mathcal{X}}'_{1:p(v')},\bar{\mathbf{a}}'_{1:p(v')-1}\right)$
and $\widehat{Z}_{\theta}(i)=\widehat{Z}'_{\theta'}$. Otherwise reject,
$\xi(i)=\xi(i-1)$ and $\widehat{Z}_{\theta}(i)=\widehat{Z}_{\theta}(i-1)$. 
\end{enumerate}
\end{enumerate}
\caption{Particle marginal Metropolis Hastings using multi-Level SMC with adaptive
level sets.\label{fig:stop_within_mcmc1} }
\end{algorithm}

Consider now that it is possible at every PMMH iteration $i$ to simulate
the auxiliary process $v$ defined upon an abstract state-space $(V,\mathscr{V})$.
Let this with associated random variable $v$, be distributed according
to $\Lambda_{\theta}$, which is assumed to possess a density with
respect to a.~$\sigma-$finite measure $dv$ written as $\overline{\Lambda}_{\theta}$.
As hinted by the notation $\Lambda_{\theta}$ should depend on $\theta$
and $v$ is meant be used to determine the sequence of levels $\{B_{n}\}_{n=1}^{p}$
for each $\theta(i)$ in PMMH. This auxiliary variable will induce
for every $\theta\in\Theta$: 
\begin{itemize}
\item a random number of level sets $p(v)\in\mathcal{J}\subset\mathbb{Z}_{+}$. 
\item a sequence of level sets $\{B_{n}(v)\}_{n=1}^{p(v)}$ with $B_{p(v)}=A$
. 
\end{itemize}
We will assume that for any $\theta\in\Theta$, Proposition \ref{prop:markov_level}
and\eqref{eq:sets} will hold $\Lambda_{\theta}-$almost everywhere,
where this time $p$ should be replaced by $p(v)$. This implies that
for every $\theta\in\Theta$ we have: 
\begin{eqnarray}
\sum_{\tau_{p(v)}\in\mathcal{I}_{p(v)}}\int_{E^{1+\tau_{p(v)}}}\overline{\gamma}_{\theta}(x_{0:\tau_{p(v)}},y,\tau_{p(v)})dx_{0:\tau_{p(v)}} & = & \sum_{\tau\in\mathcal{I}}\int_{E^{\tau+1}}\overline{\gamma}_{\theta}(x_{0:\tau},y,\tau)dx_{0:\tau},\label{eq:cond}
\end{eqnarray}
 where the expression holds $\Lambda_{\theta}-$ almost everywhere.
In Algorithm \ref{fig:stop_within_mcmc1} we propose a PMMH algorithm,
which at each step $i$ uses $\theta(i)$ to adapt the levels $\{B_{n}(v(i))\}_{n=1}^{p(v(i))}$.
For Algorithm \ref{fig:stop_within_mcmc1} we present the following
proposition that verifies varying the level sets in this way is theoretically
valid:

\begin{proposition}\label{prop:adap_stop} Assume (A\ref{assump:a5-6})
and \eqref{eq:cond} hold. Then, for any $N\geq1$: 
\begin{enumerate}
\item {The invariant density of the procedure in Algorithm \ref{fig:stop_within_mcmc1}
is defined on the space 
\[
\Theta\times V\times\bigcup_{j\in\mathcal{J}}\left(\{j\}\times\mathbb{T}_{N}^{j(N-1)+1}\times\bigg(\bigcup_{i\in\mathcal{I}_{p(j)}}\{i\}\times E^{i}\bigg){}^{N}\right)
\]
 and has the representation 
\begin{equation}
\overline{\pi}^{N}(\theta,k,v,\bar{\mathcal{X}}_{1:p(v)},\bar{\mathbf{a}}_{1:p(v)-1})=\frac{\overline{\pi}(\theta,\mathcal{X}_{1:p(v)}^{(k)})}{N^{p(v)}}\frac{\psi_{\theta}(\bar{\mathcal{X}}_{1:p(v)},\bar{\mathbf{a}}_{1:p(v)-1}))\overline{\Lambda}_{\theta}(v)}{\mathcal{M}_{1}(\mathcal{X}_{1}^{(b_{1}^{k})})\prod_{n=2}^{p(v)}\big\{\bar{W}_{n-1}^{(b_{n-1}^{k})}\mathcal{M}_{n}(\mathcal{X}_{n}^{(b_{n}^{k})}|\mathcal{X}_{n-1}^{(b_{n-1}^{k})})\big\}}\label{eq:stop_target1}
\end{equation}
 where $\overline{\pi}$ is as in \eqref{eq:target_den} and $\psi_{\theta}$
is as in \eqref{eq:stopping_density}. In addition, \eqref{eq:stop_target1}
admits $\overline{\pi}(\theta)$ as a marginal. } 
\item {The generated sequence $\left(\theta(i),\mathcal{X}_{1:p(v(i))}(i)\right){}_{i\geq0}$
satisfies 
\[
\lim_{i\rightarrow\infty}\|\mathcal{L}aw(\theta(i),\mathcal{X}_{1:p(v(i))}(i)\in\cdot)-\pi(\cdot)\|=0
\]
 where $\pi$ is as in \eqref{eq:target}.} 
\end{enumerate}
\end{proposition}

The proof is contained in the Appendix. We are essentially using an
auxiliary framework similar to \cite{pseudo-marginal}. As in \eqref{eq:target}
we included $x_{0:\tau},\tau$ in the target posterior, when we were
primarily interested in $\theta$, this time we augment the target
posterior with $v$ and the SMC variables $\bar{\mathcal{X}}_{1:p(v)},\bar{\mathbf{a}}_{1:p(v)-1}$,
which is a consequence of using PMCMC. The disadvantage is that as
the space of the posterior increases it is expected that the mixing
of the algorithm will be slower. This could be improved if we have
opted $x_{0:\tau},\tau$ and $v$ to be dependent on each other given
$\theta$, but this would need additional assumptions for the structure
of $\gamma_{\theta}$. In addition, in many applications the parameters
$v$ that determine $\{B_{n}\}_{n=1}^{p}$ appear naturally and $v$
often is low dimensional. Also, in most applications it might seem
easier to find intuition on how to construct and tune $\Lambda_{\theta}$
than computing the level sets directly from $\theta$. For example,
for the coalescent model of Section \ref{sec:coal_model} with the
mutation matrix $R$ is fixed, one can envisage for a larger value
of $\mu$ coalescent events are less likely and more level sets closer
together are needed compared to smaller values of $\mu$.

\section{Numerical Examples\label{sec:numerical}}

We will illustrate the performance of PMMH using numerical examples
on two models from population genetics. The first one deals with the
coalescent model of Section \ref{sec:coal_model} when a low dimensional
dataset is observed. This is meant as an academic/toy example suitable
for comparing different PMMH implementations. The second example is
a more realistic application and deals with a coalescent model that
allows migration of individual genetic types from one sub-group to
another \cite{bahlo,deiorio}. In both cases we will illustrate the
performance of PMMH implemented with a simple intuitive strategy for
adapting the level sets.

\subsection{The coalescent model}

We will use a known stochastic matrix $R$ with all entries equal
to $1/d$. In this example $d=4$ with and the dataset is $y=(10,5,9,5)$.
The parameter-space is set as $\Theta=[0,1.5]$ and a uniform prior
will be used. For $M_{\theta}$ we will use the optimal proposal distributions
provided by \cite{pg}. The PMMH proposal $q(\cdot|\mu(i-1))$ in
Algorithm \ref{fig:stop_within_mcmc1} is a log normal random walk,
i.e. we use $\zeta'=\zeta(i-1)+0.4\mathcal{N}(0,1)$ with $\zeta=\log(\mu)$.

We will compare PMMH when implemented with a simple adaptive scheme
for $p$ and when $p$ is fixed. In the latter case we set $p=14$.
When an adaptive strategy is employed we will sample each time $p$
directly using a multinomial distribution defined on $\{8,\dots,28\}$
with weights proportional to $\mu^{p}$. In both cases given $p$
we place the levels almost equally spaced apart.

\subsubsection{Numerical results}

\begin{figure}[ht]
\centering \includegraphics[width=0.25\textwidth]{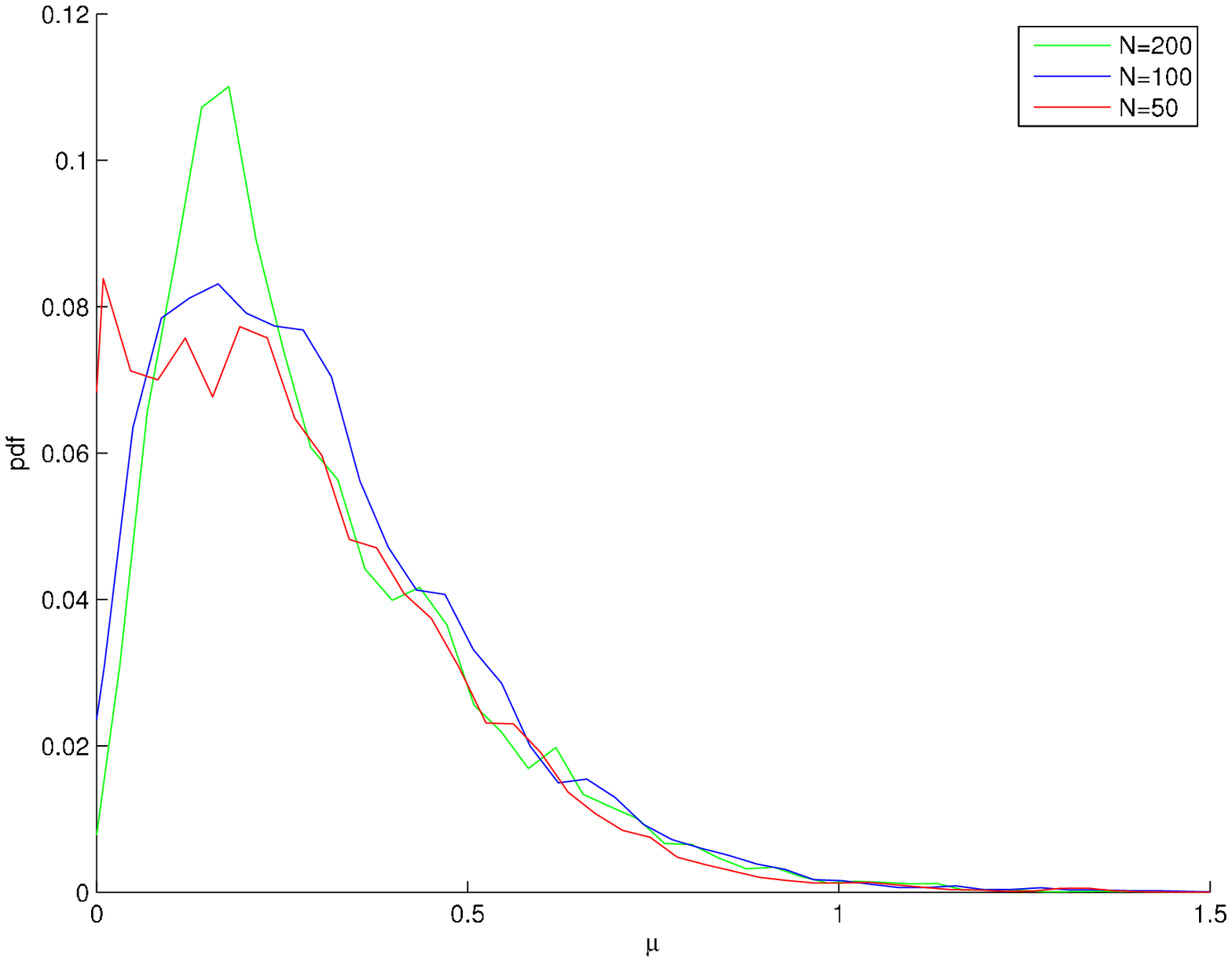}\includegraphics[width=0.25\textwidth]{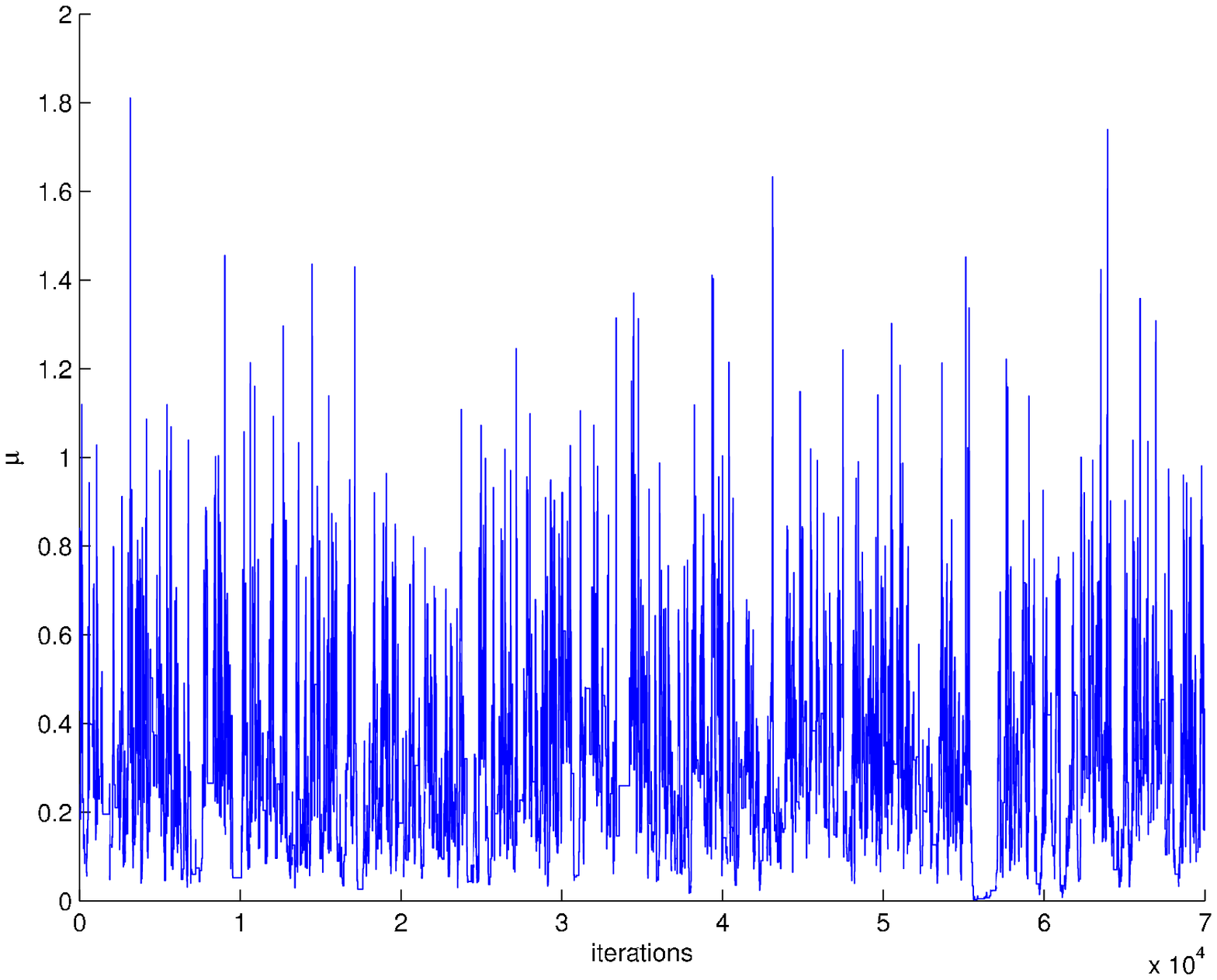}\includegraphics[width=0.25\textwidth]{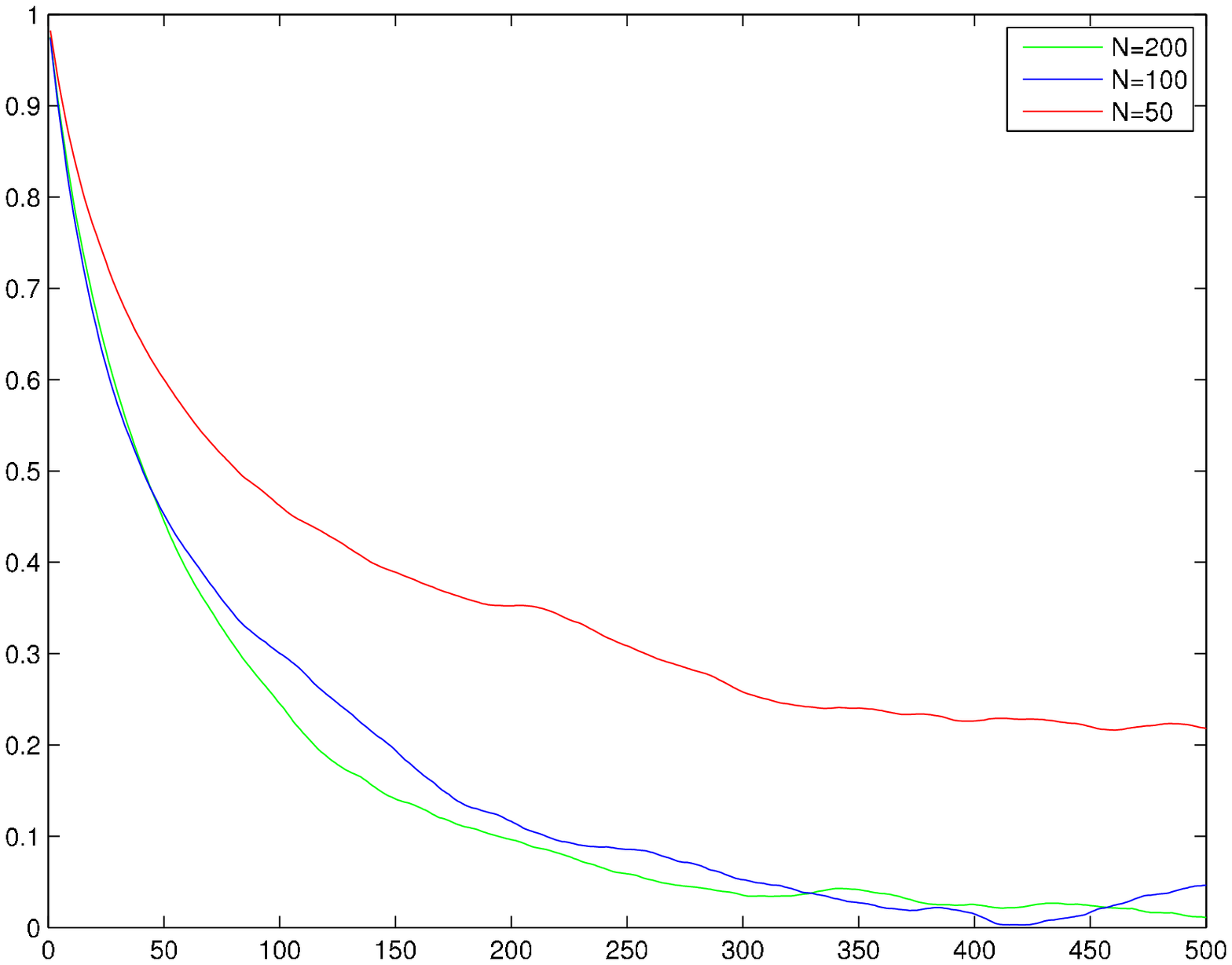}
\caption{PMMH for the coalescent without adaptation. A fixed number of $14$
level sets is used. Left: estimated pdf of $\mu$ for $N=50,100,200$.
Centre: the trace plot for $N=100$. Right: autocorrelation plots
for $N=50,100,200$. The average acceptance ratio was $0.07$, $0.08$
and $0.10$ respectively.}
 \label{fig:plots_noadap} 
\end{figure}

\begin{figure}[ht]
\centering \includegraphics[width=0.25\textwidth]{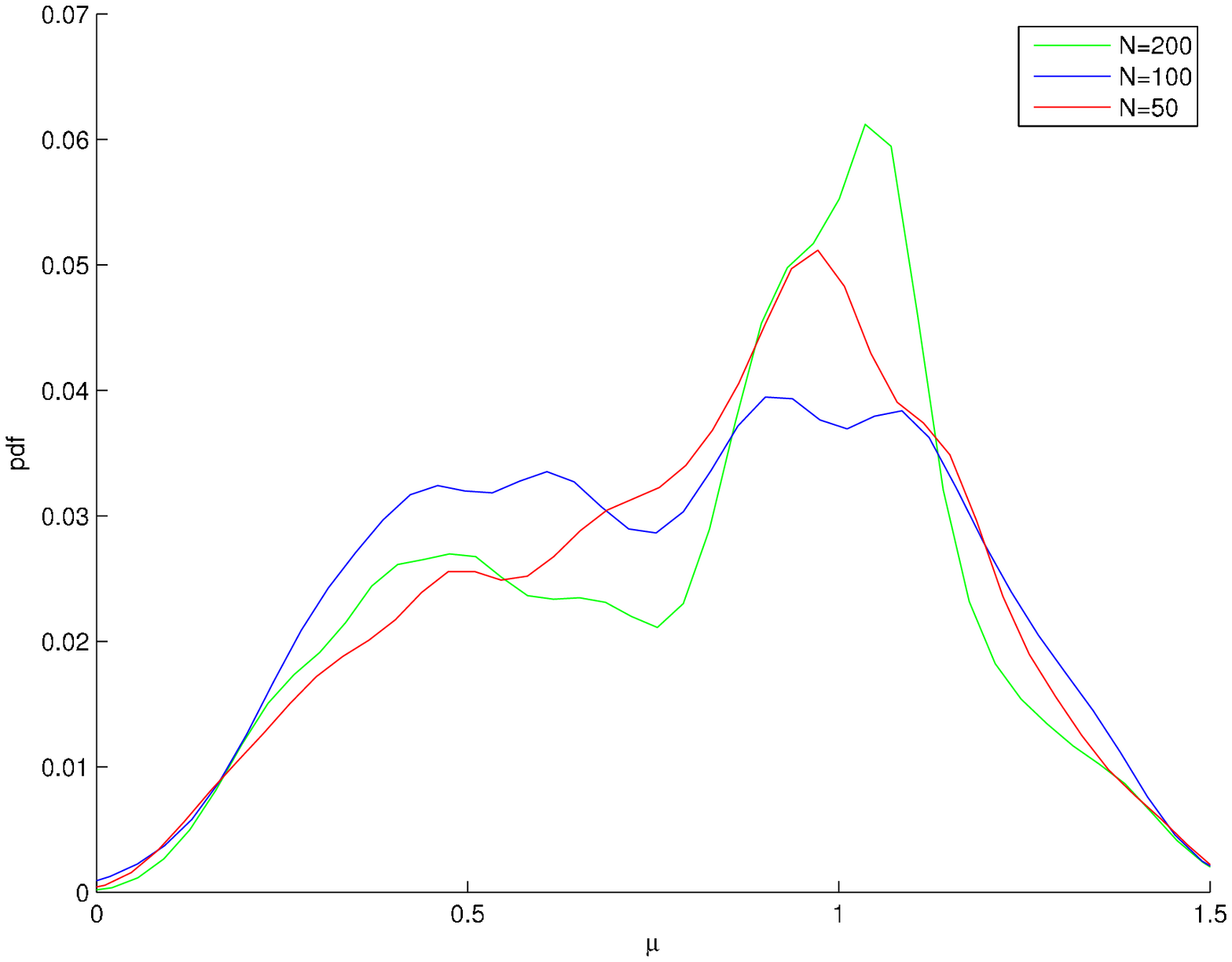}\includegraphics[width=0.25\textwidth]{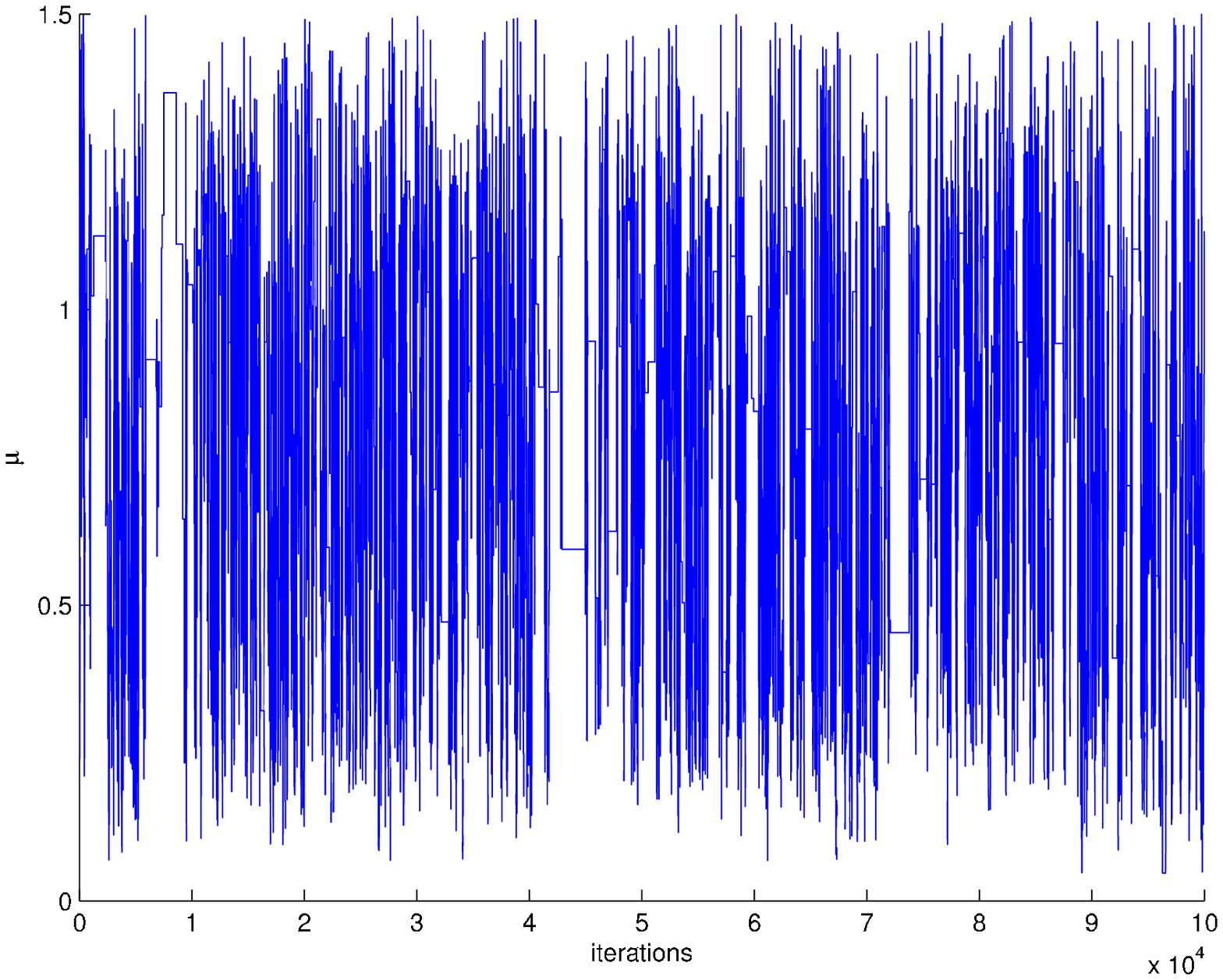}\includegraphics[width=0.25\textwidth]{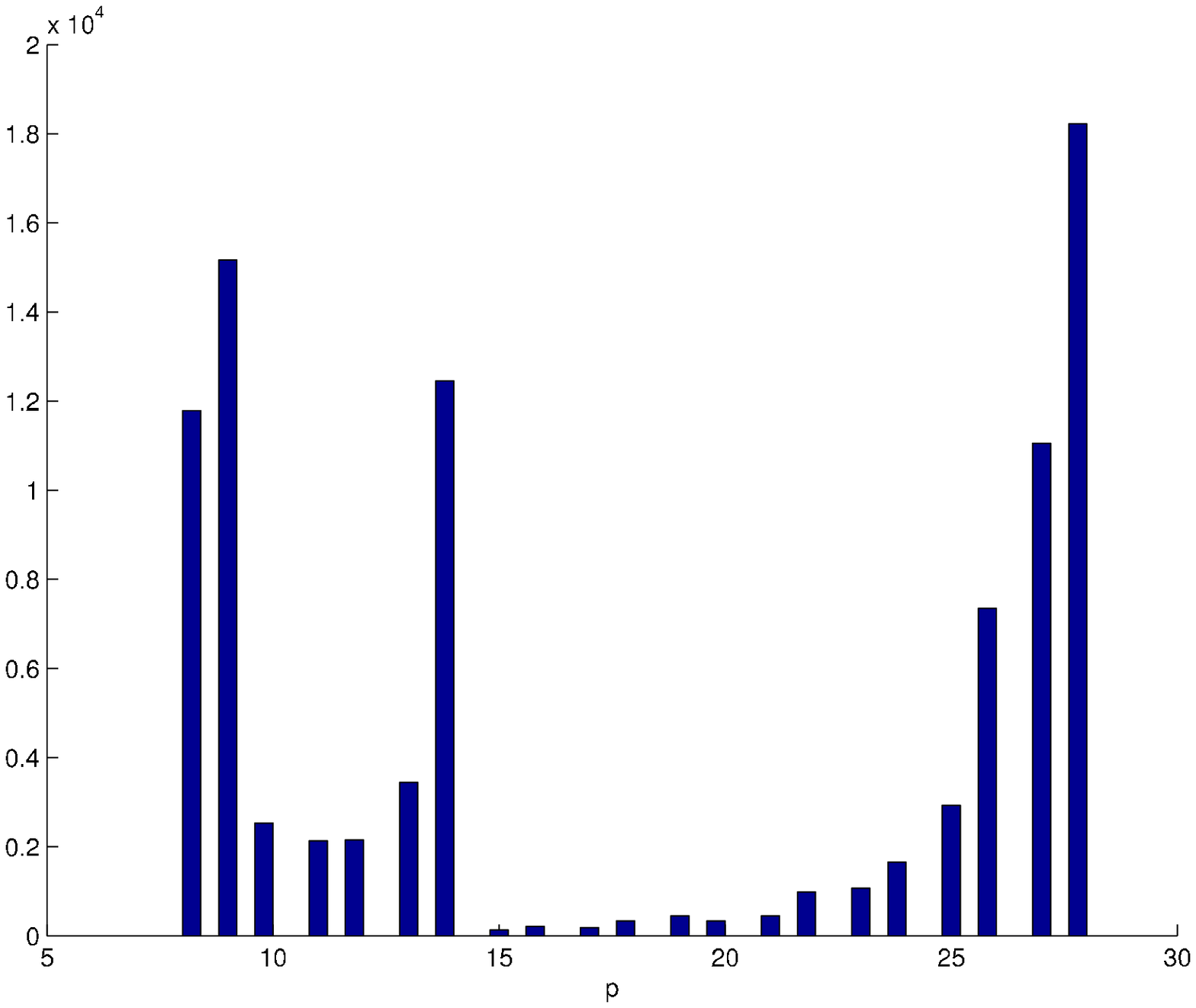}\includegraphics[width=0.25\textwidth]{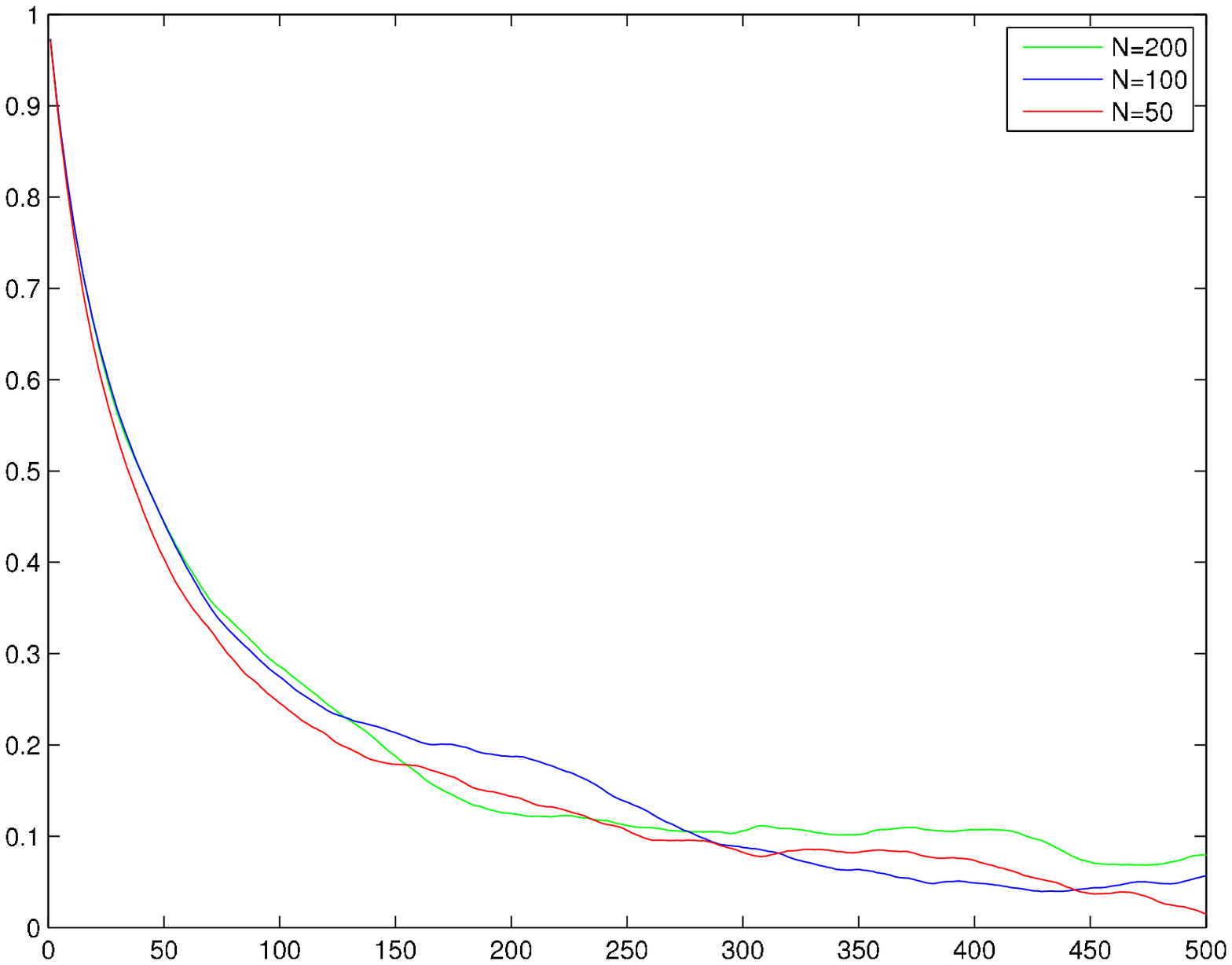}
\caption{PMMH for the coalescent adaptation. The number of levels sampled is
proportional to $\mu^{p}$. Far left: estimated pdf of $\mu$ for
$N=50,100,200$. Central left: the trace plot for $N=100$. Central right:
histogram of number of levels in the posterior for $N=100$. Far
right: autocorrelation function plots for $N=50,100,200$. The average
acceptance ratio was $0.10$, $0.11$ and $0.13$ respectively.}
\label{fig:plots_adap} 
\end{figure}

The adaptive and normal versions were run with $N=50,100,200$ for
$10^{5}$ iterations. In each case the algorithm took approximately
$2.5$, $5$, $10$ hours to complete when implemented in Matlab and
run on a Linux workstation using a Intel Core 2 Quad Q9550 CPU at
2.83 GHz. The results are shown in Figure \ref{fig:plots_noadap}
and \ref{fig:plots_adap}. We observed that when we varied the number
of levels, this allowed the sampler to traverse through a bigger part
of the state space compared to when a fixed number of levels is used.
As a result the estimated pdf of the adaptive case manages to include
a second mode that is not seen in the non adaptive case. 
In the fixed levels case we see a clear improvement with increasing
$N$, although the difference in the mixing between $N=100$ and $200$
is marginal. In the adaptive case the sampler performed well even
with lower values of $N$.


\subsection{The coalescent model with migration}

\label{sec:migration}

The model is similar to the one as described in Section \ref{sec:coal_model}.
The major difference is that this time genetic types are of classified
into sub-groups within which most activity happens. In addition, individuals
are allowed to migrate from one group to another. We commence with
a brief description of the model and refer the interested reader to
\cite{bahlo,deiorio} for more details. As in Section \ref{sec:coal_model}
we will consider the process forward in time. Let $g$ be the number
of groups and the state at time $t$ be composed as the concatenation
of $g$ groups of different genetic types as: 
\[
x_{t}=(x_{1,t}^{1},\dots,x_{1,t}^{d},\dots,x_{g,t}^{1},\dots,x_{g,t}^{d})
\]
 The process under-goes split, mutation and migration transitions
as follows: 
\begin{eqnarray*}
X_{j} & = & X_{j-1}+e_{\alpha,i}\\
X_{j} & = & X_{j-1}-e_{\alpha,i}+e_{\alpha,l}\\
X_{j} & = & X_{j-1}-e_{\alpha,i}+e_{\beta,i},
\end{eqnarray*}
 where $\alpha,\beta\in\{1,\dots,g\}$ with $\alpha\neq\beta$ and
$e_{\alpha,i}$ is a vector with a zero in every element except the
$(\alpha-1)g+i$ -th one. Similarly to the simpler model of Section
\ref{sec:coal_model} the transition probabilities are parameterised
by the mutation parameter $\mu$, mutation matrix $R$ and a migration
matrix $G$. The latter is a symmetric matrix with zero values on
the diagonal and positive values on the off-diagonals. Finally the
data is generated when at time $\tau$ the number of individuals in
the population reaches $m$, and $y=y^{1:gd}=x_{\tau}$.

As for the model described in Section \ref{sec:coal_model} one can
reverse time and employ an backward sampling forward weighting importance
sampling method; see \cite{deiorio} for the particular implementation
details. In our example we generated data with $m=100$, $d=64$ and
$g=3$. This is quite a challenging set-up. As in the previous example
we set the mutation matrix $R$ to be known and uniform and we will
concentrate on inferring the $\theta=(\mu,G)$. Independent gamma
priors with shape and scale parameters equal to $1$ were adopted
for each of the parameters.

\subsubsection{Numerical results}

We implemented PMMH using $N=50,100,200$ and a simple adaptive scheme
for $p$. We allow $p\in\left\{ 10,20,33\right\} $ and use approximately
equal spacing between the levels. We choose each $p$ with probability
proportional to $p{}^{\log\{\mu+\sum_{i>j}G_{ij}+1\}}.$ The proposals
for the parameters were Gaussian random-walks on the log-scale. The
algorithm was implemented in C/C++ was run for $10^{5}$ iterations,
which took approximately 3, 6 and 12 hours to complete. Whilst the
run-time is quite long it can be improved by at least one order of
magnitude if the SMC is implemented on Graphical Processing Units
(GPU) as in \cite{leegpu}.

\begin{figure}[ht]
\centering \includegraphics[width=0.25\textwidth]{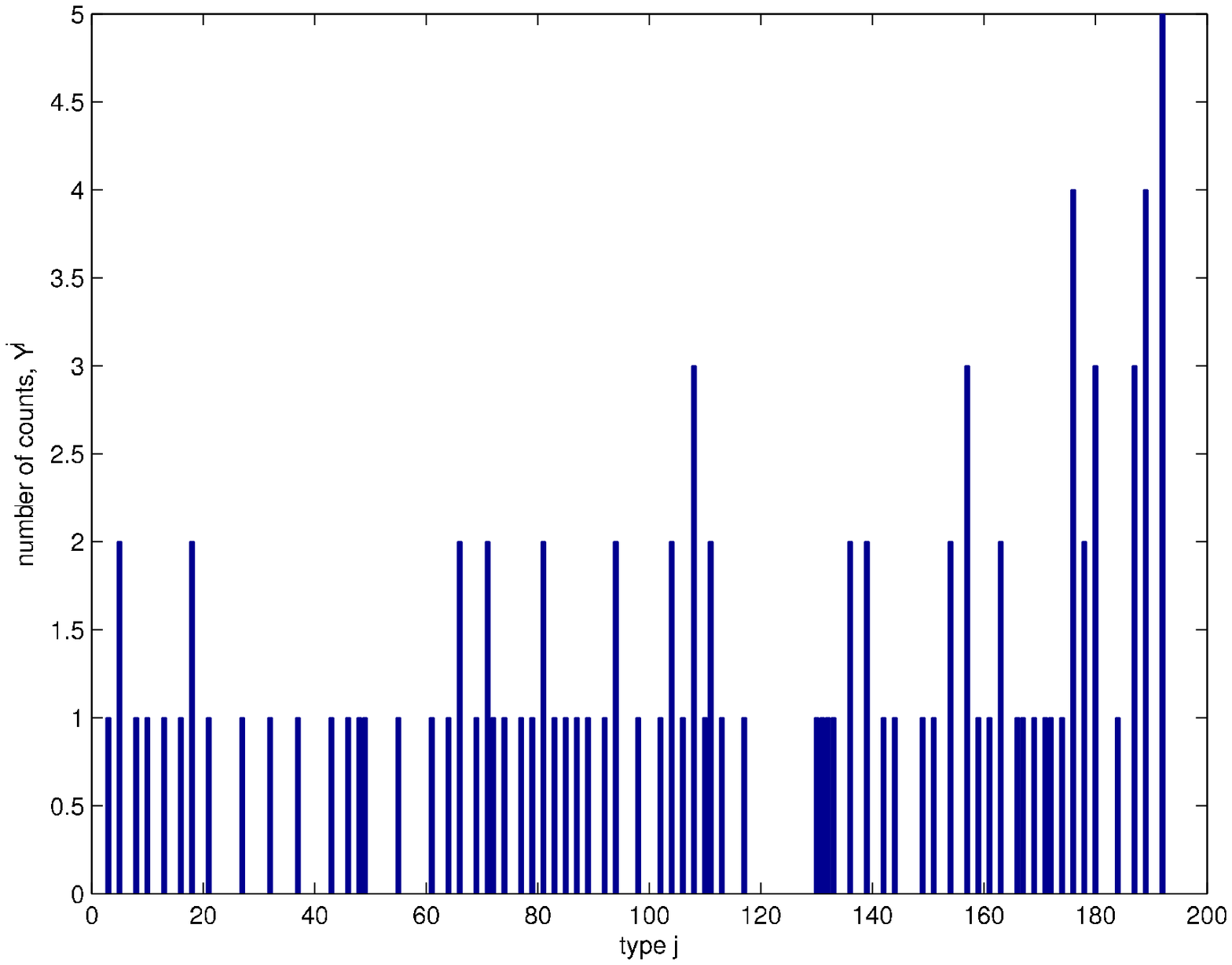}\includegraphics[width=0.25\textwidth]{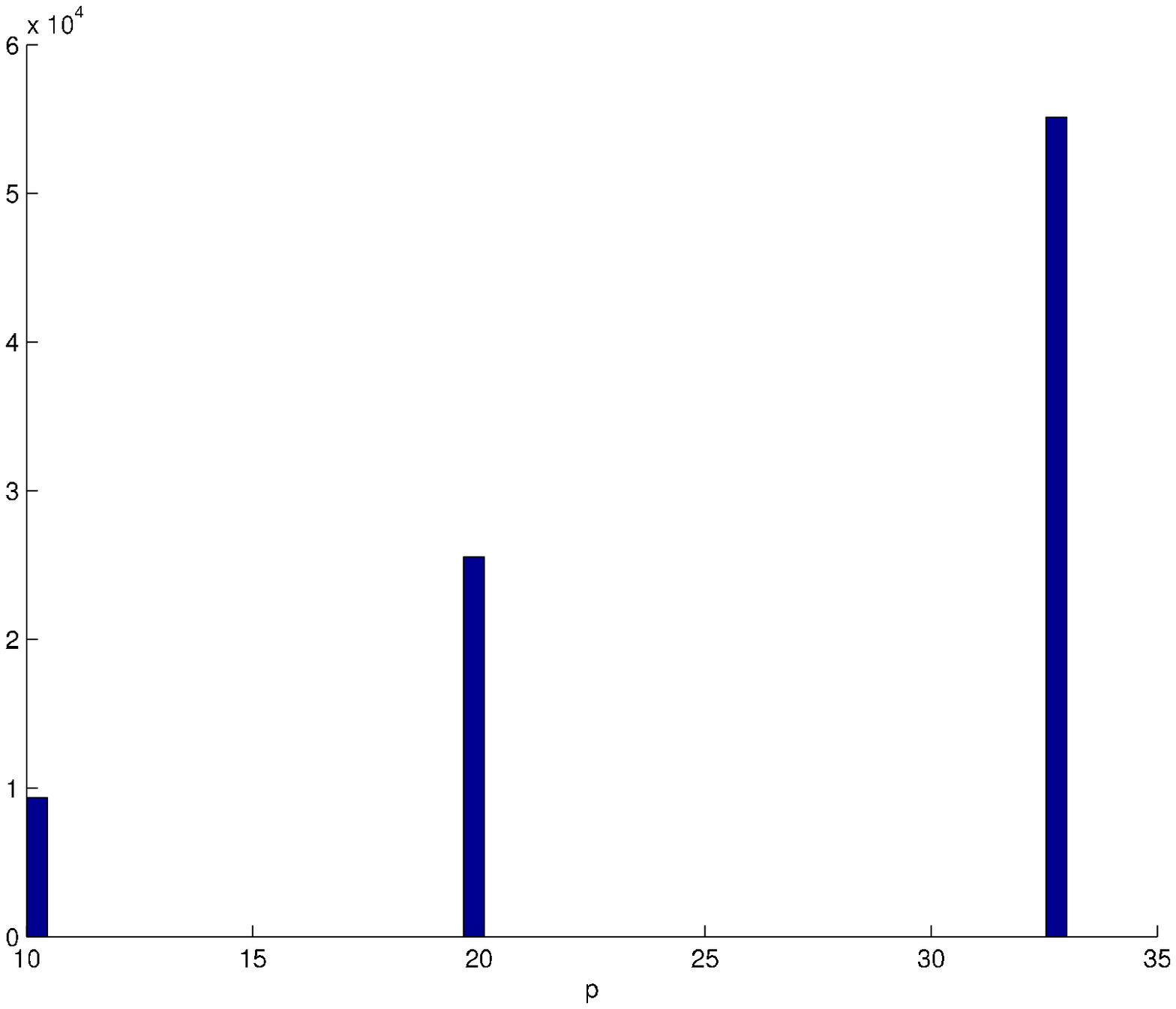}
\caption{Left: Dataset for the Coalescent with Migration. Right: Histogram
of number of levels $p$ in the resulting posterior for $N=100$.}

\label{fig:plots2} 
\end{figure}

\begin{figure}[ht]
\includegraphics[width=0.25\textwidth]{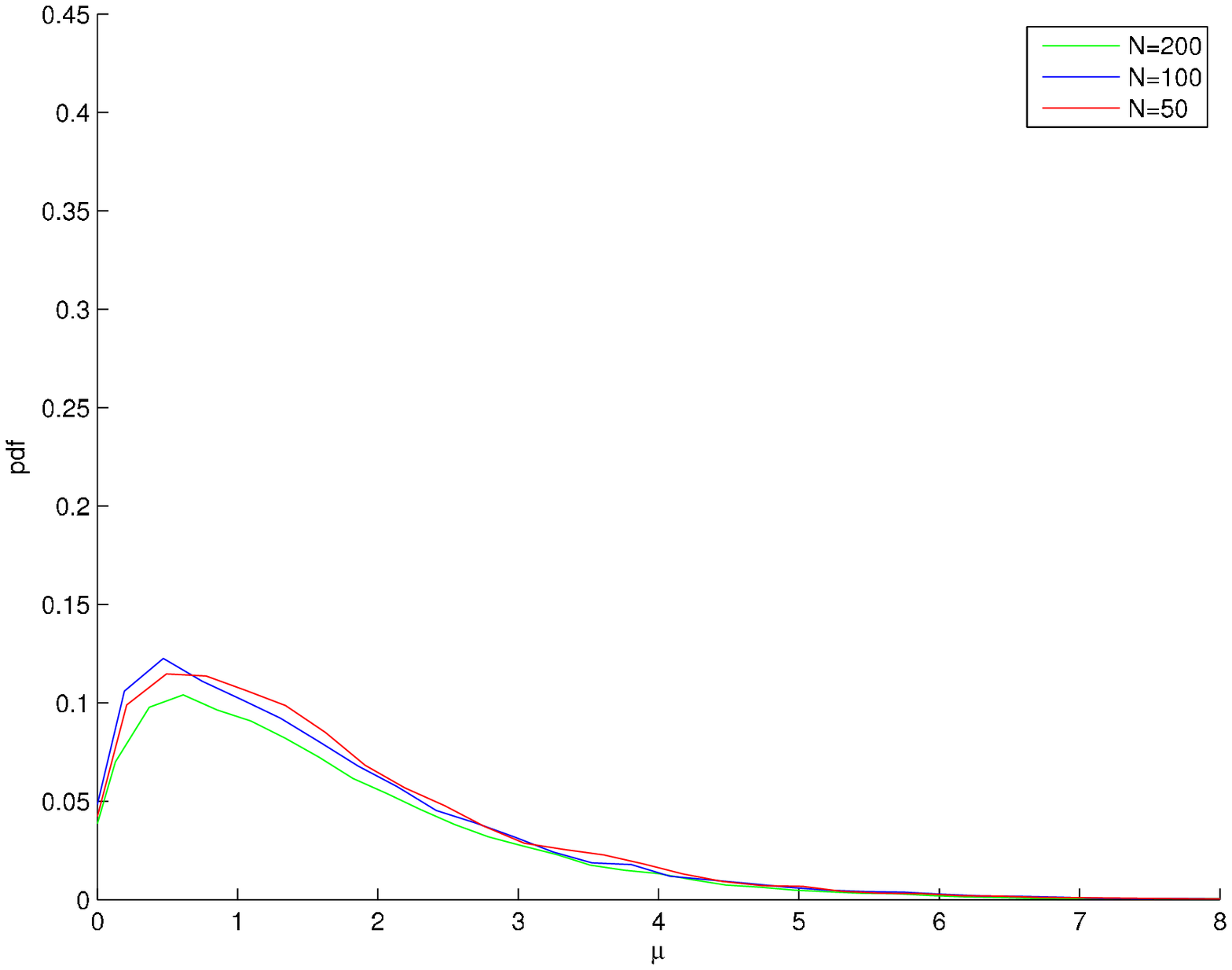}\includegraphics[width=0.25\textwidth]{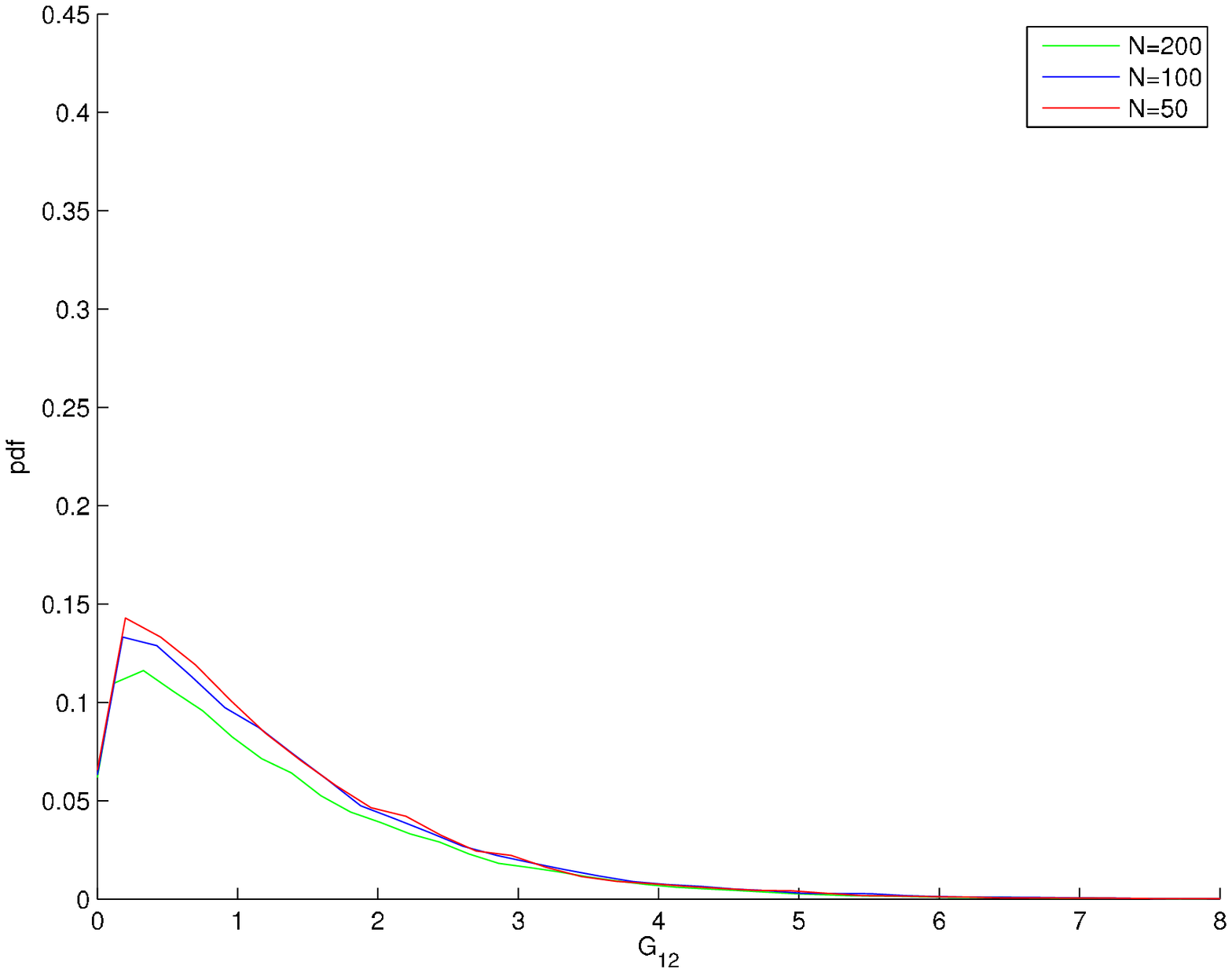}\includegraphics[width=0.25\textwidth]{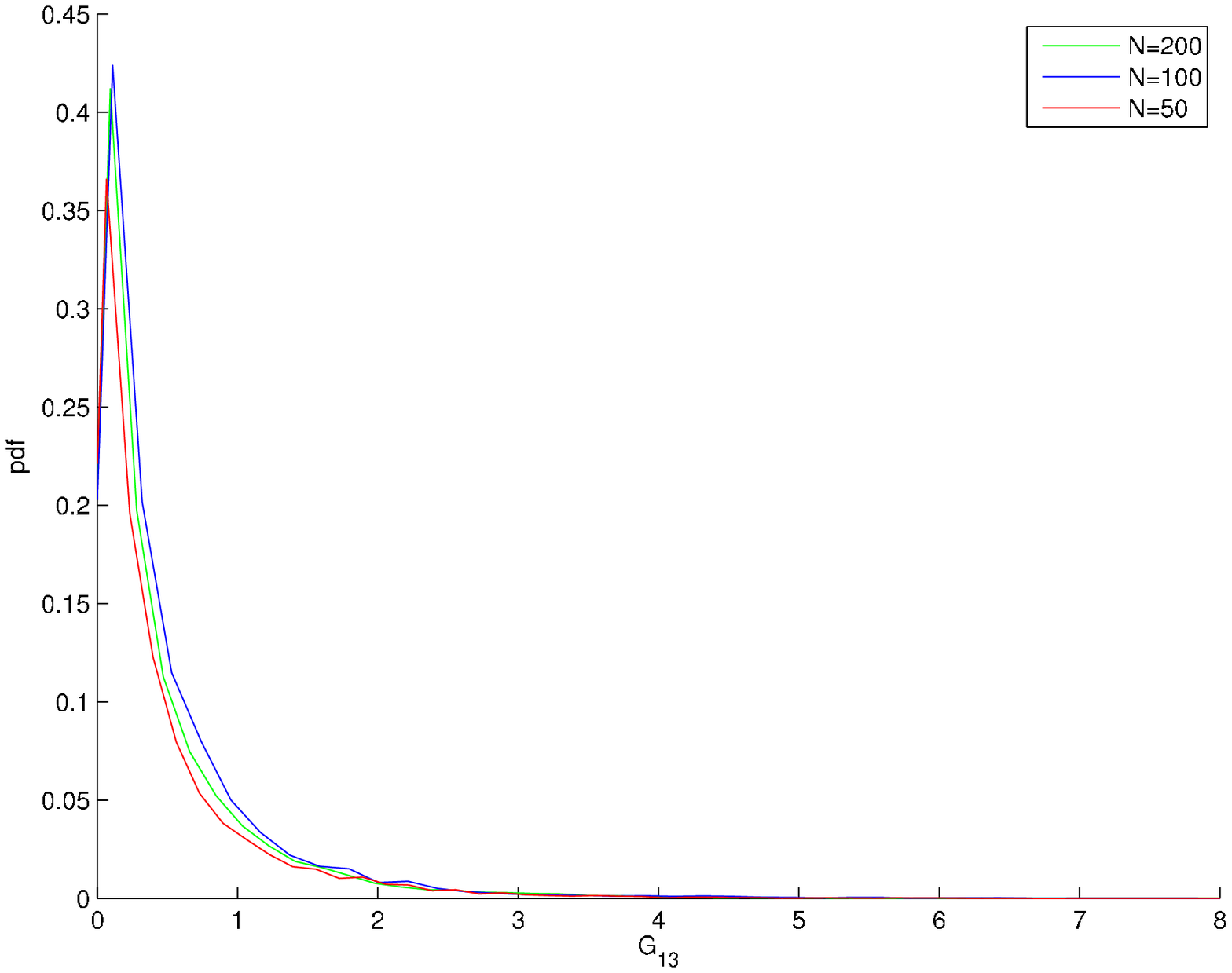}\includegraphics[width=0.25\textwidth]{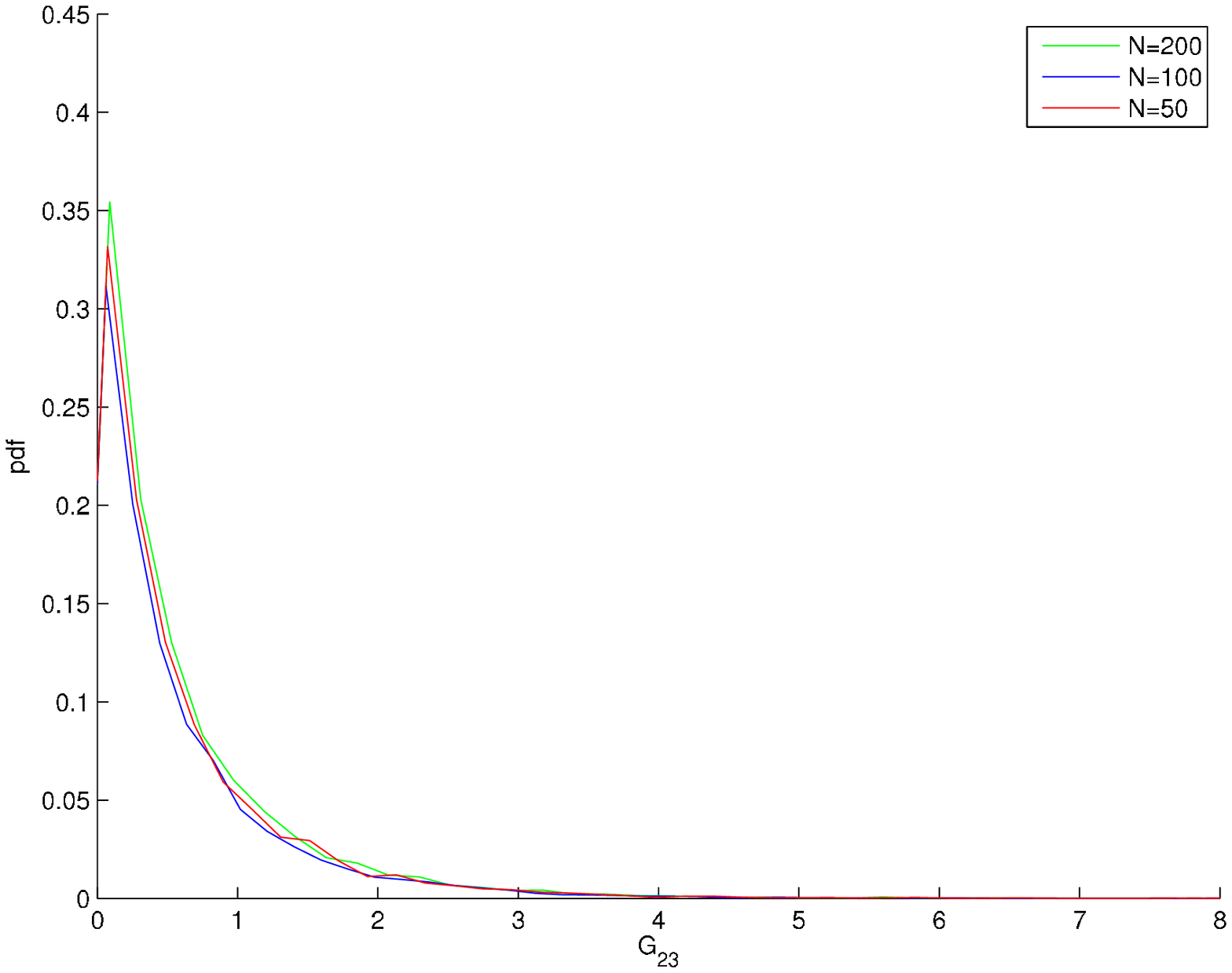}\\
\includegraphics[width=0.25\textwidth]{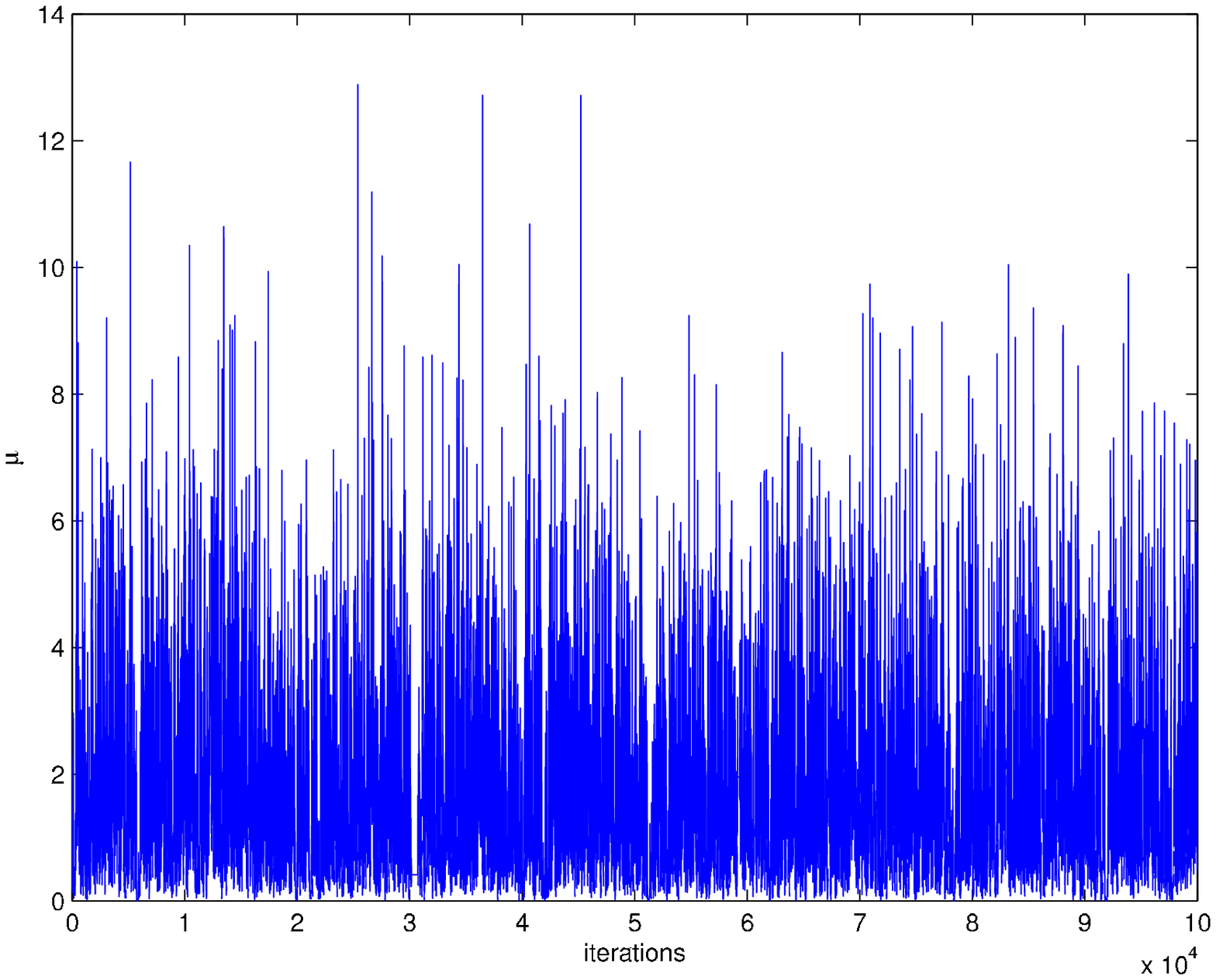}\includegraphics[width=0.25\textwidth]{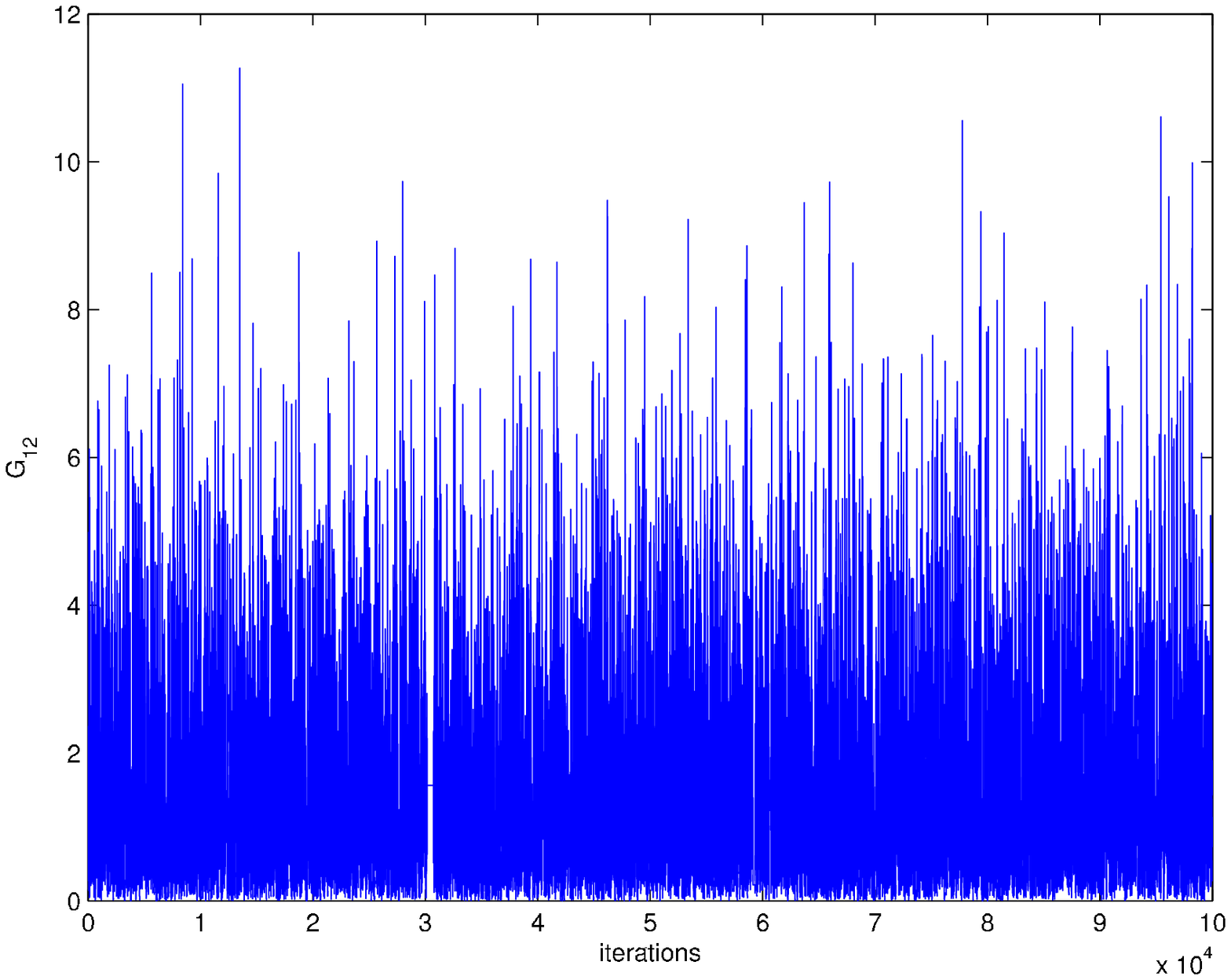}\includegraphics[width=0.25\textwidth]{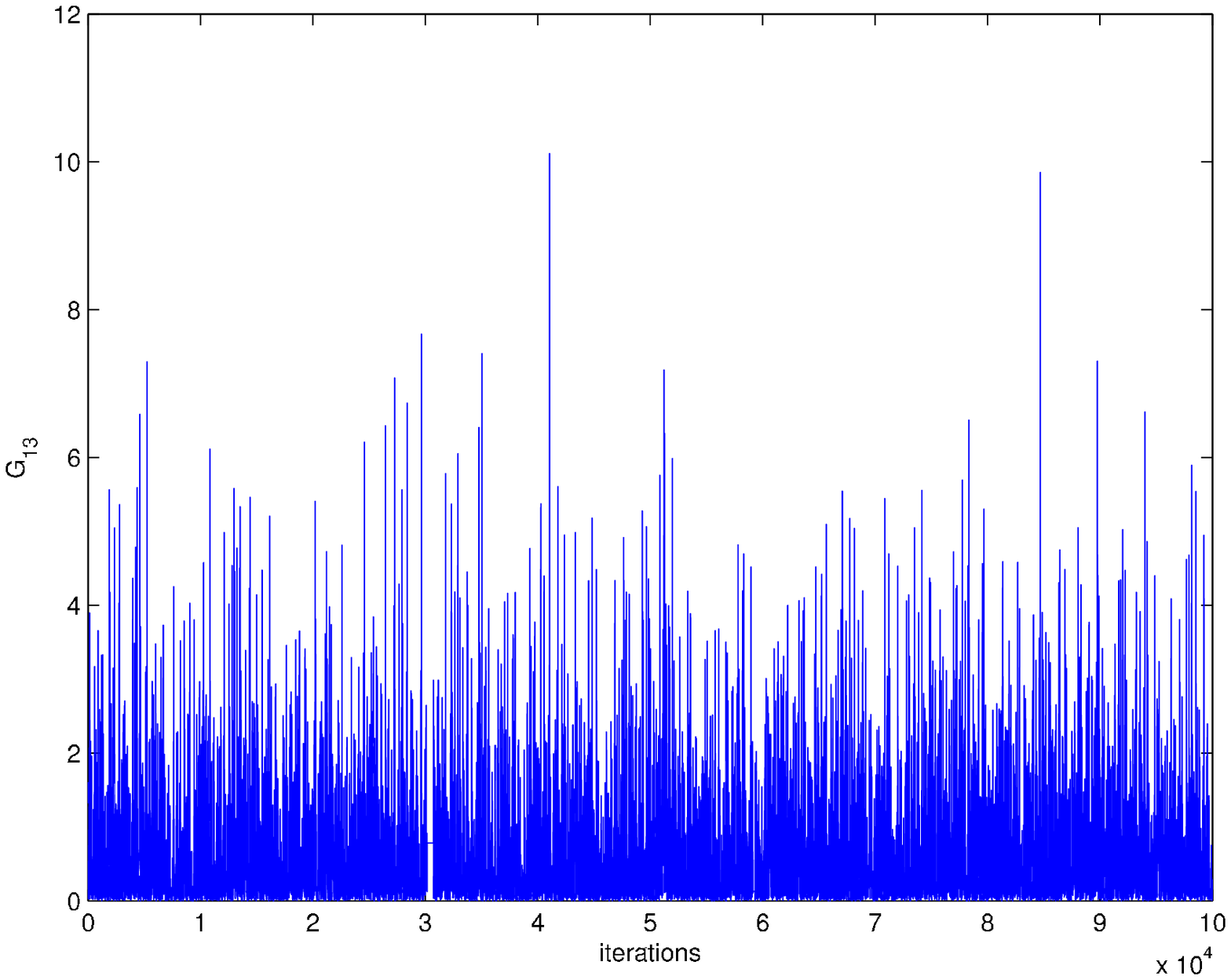}\includegraphics[width=0.25\textwidth]{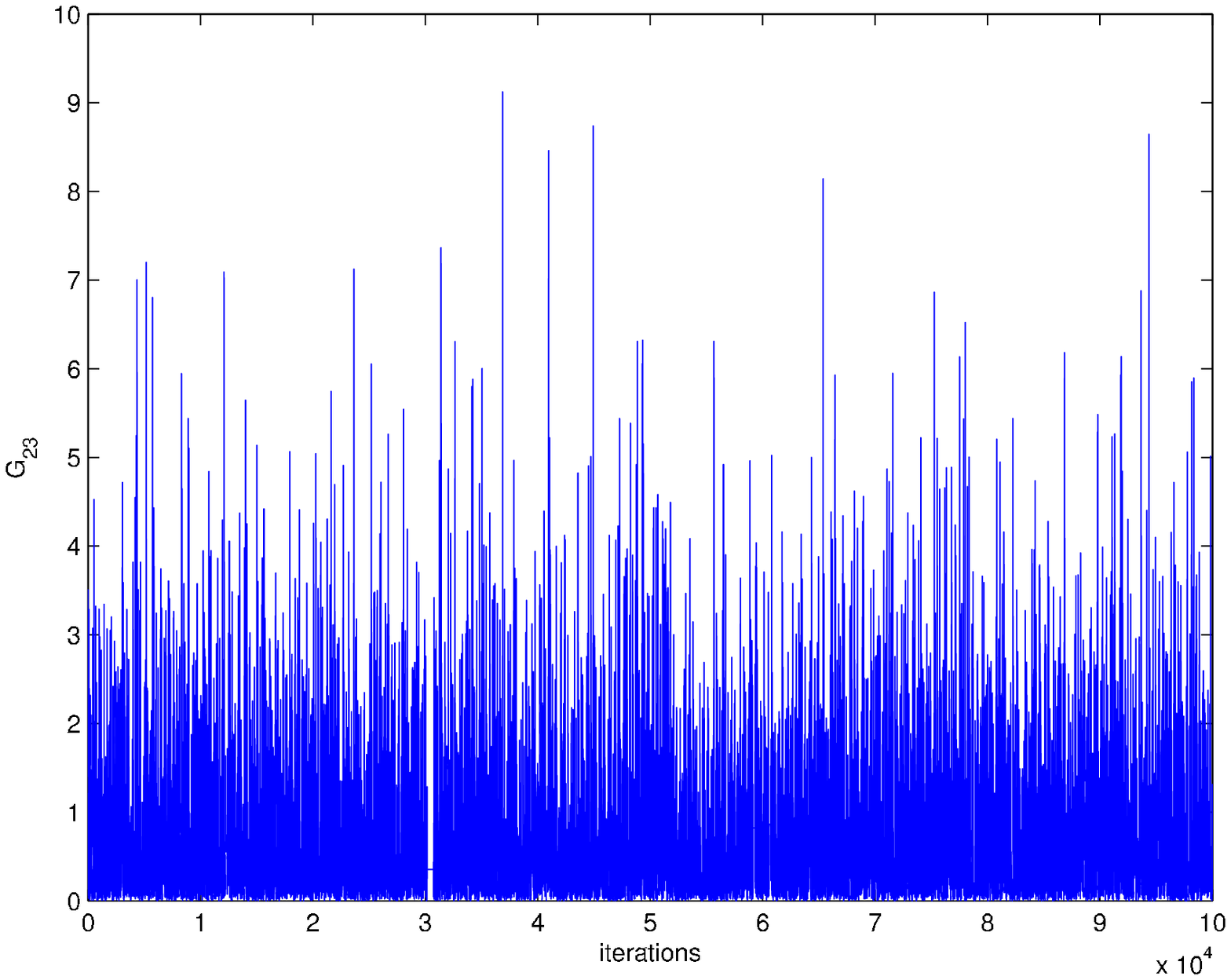}\\
\includegraphics[width=0.25\textwidth]{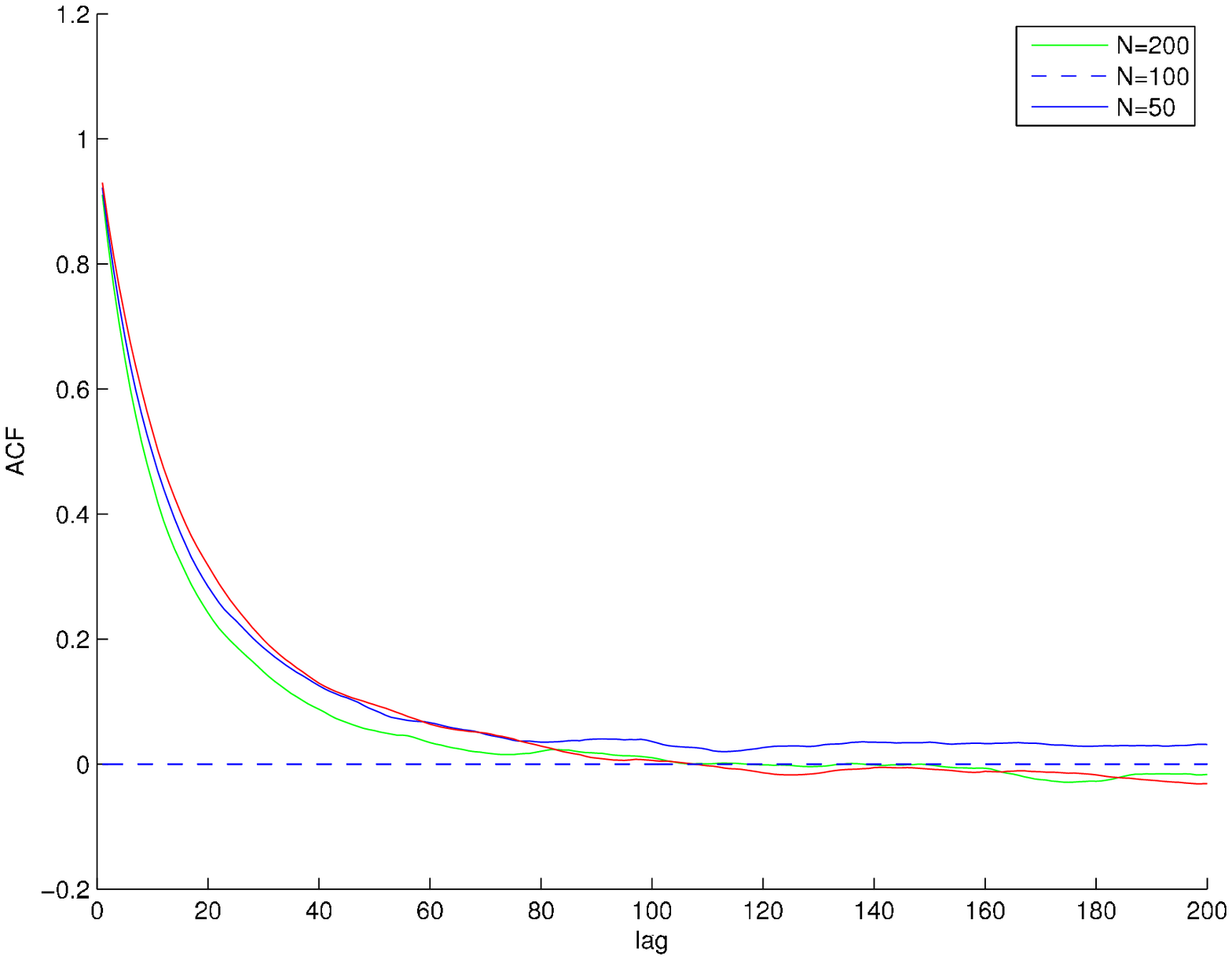}\includegraphics[width=0.25\textwidth]{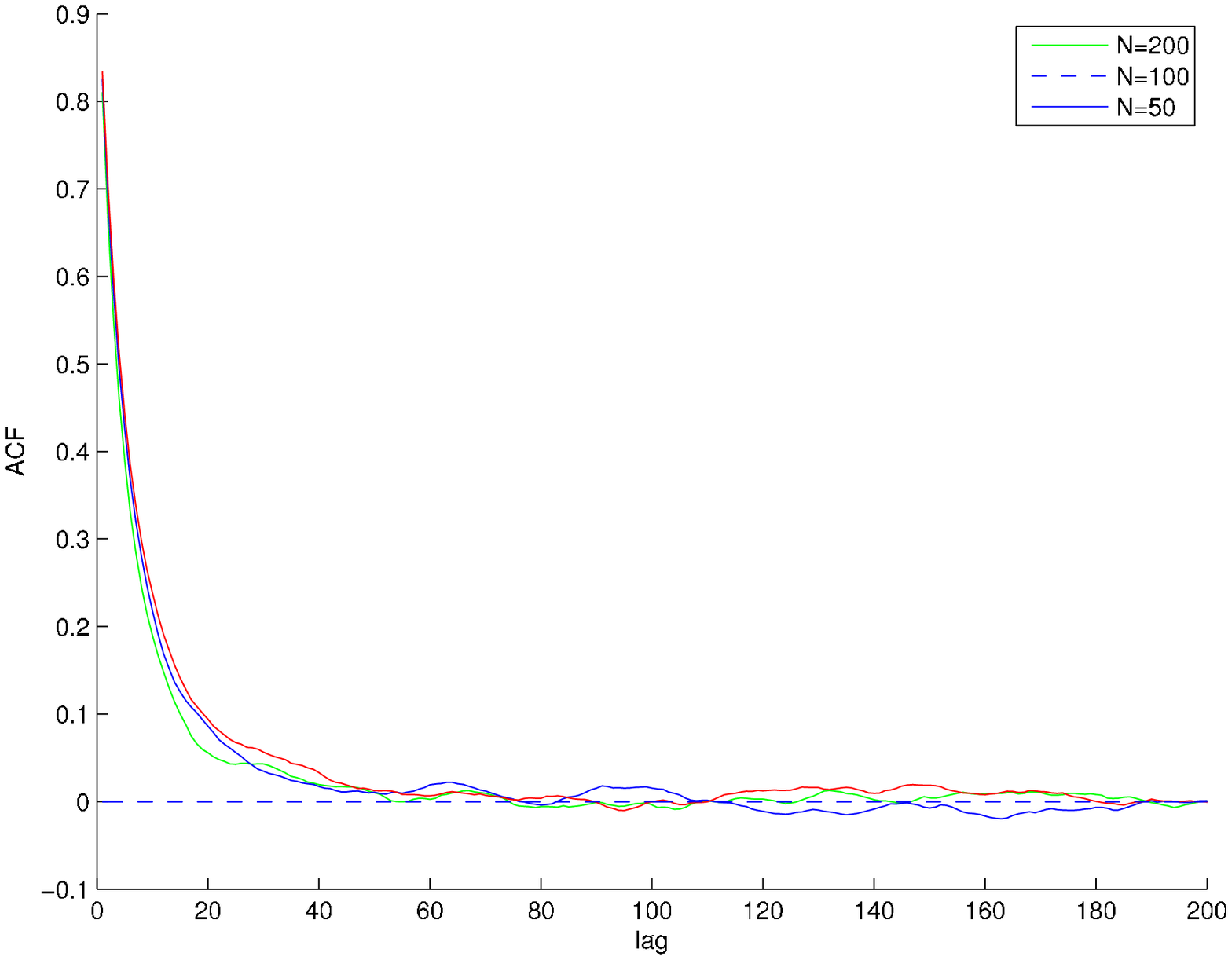}\includegraphics[width=0.25\textwidth]{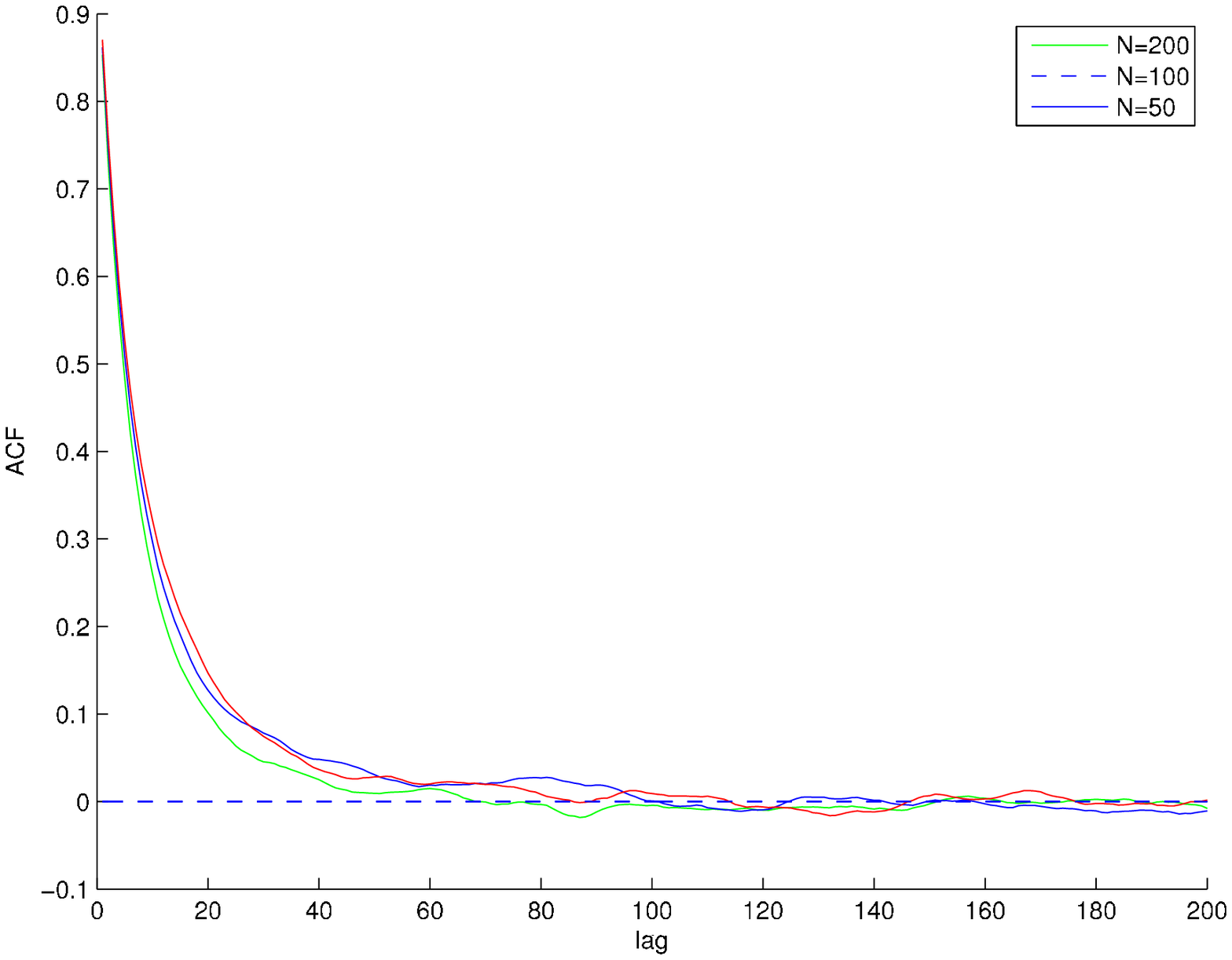}\includegraphics[width=0.25\textwidth]{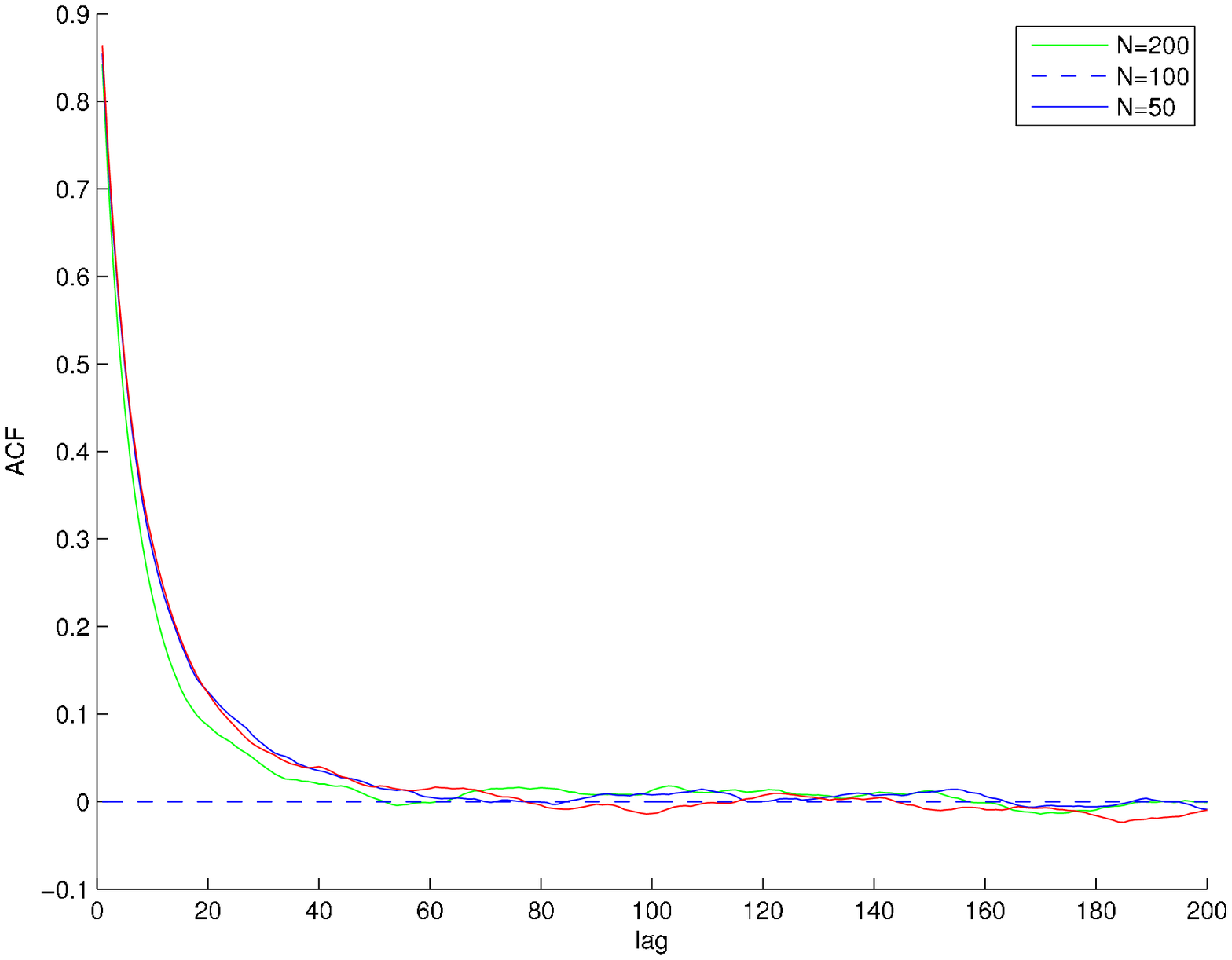}

\caption{PMMH for for the Coalescent with Migration for $N=50,100,200$. Top
row: estimated pdfs for $\mu$, $G_{12}$, $G_{13}$, $G_{23}$ (from
left to right). Middle: trace plots for $N=100$. Bottom: autocorrelation
function plots. The acceptance rate was $0.34,0.37,0.4$ respectively.}

\label{fig:plots1} 
\end{figure}

For the dataset plotted in Figure \ref{fig:plots2} (left) the results
are plotted in Figures \ref{fig:plots2} (right) and \ref{fig:plots1}.
The auto-correlation and trace plots indicate that the sampler mixes
reasonably well for every $N$. These results in this example are
encouraging as to the best of our knowledge Bayesian inference has
not been attempted for this class of problems. We expect that practitioners
with insight in the field of population genetics can come with more
sophisticated MCMC proposals or adaptive schemes for the level sets,
so that the methodology can be extended to realistic applications.

\section{Discussion}

\label{sec:summary}

In this article we have presented a multi-level PMCMC algorithm which
allows one to perform Bayesian inference for the parameters of a latent
stopped processes. In terms of methodology the main novelty of the
approach is that uses auxiliary variables to adaptively compute the
level sets with $\theta$. The general structure of this auxiliary
variable allows it to incorporate the use of independent SMC runs
with less particles to set the levels. In the numerical examples we
demonstrated that the addition auxiliary variables slow down the convergence
of PMCMC, but this seemed a reasonable compromise in terms of performance
compared when fixed number of level sets were used. The proposed algorithm
requires considerable amount of computation, but to the authors best
knowledge for such problems there seems to be a lack of alternative
approaches. Also, recent developments GPU hardware can be adopted
to speed up the computations even by orders of magnitude as in \cite{leegpu}.

There are several extensions to the work here, which may be considered.
Firstly, the scheme that is used to adapt the level sets relies mainly
on intuition. We found simple adaptive implementations to work well
in practice. In the rare events literature one may find more systematic
techniques to design the level sets, based upon optimal control \cite{dean}
or simulation \cite{cerou_static}. Although these methods are not
examined here, they can be characterised using alternative auxiliary
variables similar to the ones in Proposition \ref{prop:adap_stop},
so the auxiliary variable framework we use is quite generic. In addition,
we emphasise that within a PMCMC framework one may also include multi-level
splitting algorithms instead of SMC, which might appeal practitioners
familiar with multi-level splitting.

Secondly, one could seek to use these ideas within a SMC sampler framework
of \cite{delm:06} as done in \cite{chopin_jacob}. As noted in the
latter article, a sequential formulation can improve the sampling
scheme, sometimes at a computational complexity that is the same as
the original PMCMC algorithm. In addition, this article focuses on
the PMMH algorithm, so clearly extensions using particle Gibbs and
block updates might prove valuable for many applications.

Finally, from a modelling perspective, it may be of interest to apply
our methodology in the context of hidden Markov models. In this context,
one has 
\[
\xi(y|x_{0:\tau})=\prod_{i=0}^{\tau}g_{\theta}(y_{i}|x_{i})
\]
 with $g_{\theta}(\cdot|x)$ being the conditional likelihood of the
observations. It would be important to understand, given a range of
real applications, the feasibility of statistical inference, combined
with the development of our methodology. An investigation of the effectiveness
of such a scheme when applied to queuing networks is currently underway.

\subsubsection*{Acknowledgement}

We thank Arnaud Doucet and Maria De Iorio for many conversations on
this work. The first author was supported by a Ministry of Education
grant. The second author was kindly supported by the EPSRC programme
grant on Control For Energy and Sustainability EP/G066477/1. Much
of this work was completed when the first author was employed by the
Department of Mathematics, Imperial College London.

\section*{Appendix}

\begin{proof}{[}Proof of Proposition \ref{prop:conv_rate}{]} The
result is a straight forward application of Theorem 6 of \cite{roberts2}
which adapted to our notation states:

\[
\|\mathcal{L}aw(\mathcal{X}_{1:p}(i)\in\cdot|\xi(0))-\check{\pi}_{\theta}(\cdot)\|\leq\mathbb{E}_{\pi_{p}^{N}}\bigg[\bigg(1-\left(\mathbb{E}_{\psi_{\theta}}\left[1\wedge\frac{\hat{Z}_{p}(\Xi)}{\hat{Z}_{p}(\xi(0))}\bigg|\xi(0)\right]\wedge\mathbb{E}_{\psi_{\theta}}\left[1\wedge\frac{\hat{Z}_{p}(\Xi)}{\hat{Z}_{p}(\xi)}\bigg|\xi\right]\right)\bigg)^{i}\bigg],
\]
 where the conditional expectation is the expectation w.r.t.~the
SMC algorithm (i.e.~$\Xi\sim\psi_{\theta}$) and the outer expectation
is w.r.t.~the PIMH target (i.e.~$\xi\sim\pi_{p}^{N}$). We also
denote the estimate of the normalizing constant as $\hat{Z}_{p}(\cdot)$
with $\cdot$ denoting which random variables generate the estimate.

Now, clearly via (A\ref{assump:a1}) 
\[
w_{n}(X_{0:\tau_{n}})\leq\frac{\rho^{\tau_{n}}}{\rho^{\tau_{n-1}}\varphi^{\tau_{n}-\tau_{n-1}}}\leq\bigg[\frac{1}{\rho\varphi}\bigg]^{\tau_{n}+\tau_{n-1}}
\]
 with the convention that $\tau_{0}=0$. Thus, it follows that 
\[
\prod_{n=1}^{p}\frac{1}{N}\sum_{j=1}^{N}W_{n}^{(j)}\leq\prod_{n=1}^{p}\bigg[\frac{1}{\rho\varphi}\bigg]^{\bar{\tau}_{n}+\bar{\tau}_{n-1}}\leq\bigg[\frac{1}{\rho\varphi}\bigg]^{2\sum_{n=1}^{p}\bar{\tau}_{n}}
\]
 and we obtain: 
\[
\frac{Z_{p}(\Xi)}{\hat{Z}_{p}(\cdot)}\geq Z_{p}(\Xi)\left(\rho\varphi\right)^{2\sum_{n=1}^{p}\bar{\tau}_{n}}.
\]
 Note that by assumption $Z_{p}(\Xi)\left(\rho\varphi\right)^{2\sum_{n=1}^{p}\bar{\tau}_{n}}\leq1$
and thus we have 
\[
\|\mathcal{L}aw(\mathcal{X}_{1:p}(i)\in\cdot|\xi(0))-\check{\pi}_{\theta}(\cdot)\|\leq\bigg(1-\mathbb{E}_{\psi_{\theta}}[Z_{p}(\Xi)\left(\rho\varphi\right)^{2\sum_{n=1}^{p}\bar{\tau}_{n}}]\bigg)^{i}
\]
 Given \cite[Theorem 7.4.2,  Equation (7.17), page 239]{delmoral}
and the fact that $\gamma_{\theta}$ is defined to be strictly positive
in (A\ref{assump:a1}) we have that the SMC approximation $\hat{Z}_{p}(\cdot)$
is an unbiased estimate of the normalizing constant $Z_{p}$ 
\begin{equation}
\mathbb{E}_{\psi_{\theta}}[Z_{p}(\Xi)]=Z_{p},\label{eq:SMC_unbiased}
\end{equation}
 and we can easily conclude. 
\end{proof}

\begin{proof}{[}Proof of Proposition \ref{prop:stop_within_mcmc}{]}
The proof of parts 1. and 2. follows the line of arguments used in
Theorem 4 of \cite{pmcmc}, which we will adapt to our set-up. The
main difference lies in the multi-level construction and second statement
regarding the marginal of $\overline{\pi}^{N}$. For the validity
of the multi-level set-up we will rely on Proposition \ref{prop:markov_level}.

Suppose we design a Metropolis Hastings kernel with invariant density
$\overline{\pi}^{N}$ and use a proposal $q^{N}(\theta,k,\bar{\mathcal{X}}_{1:p},\bar{\mathbf{a}}_{1:p-1})=\psi_{\theta}(\bar{\mathcal{X}}_{1:p},\bar{\mathbf{a}}_{1:p-1})f(k|W_{p})\bar{q}(\theta(i-1)|\theta^{\prime})=\psi_{\theta}(\bar{\mathcal{X}}_{1:p},\bar{\mathbf{a}}_{1:p-1})\bar{W}_{p}^{(k)}\bar{q}(\theta|\theta^{\prime})$
. Then 
\begin{align*}
\frac{\bar{\pi}_{p}^{N}(\theta,k,\bar{\mathcal{X}}_{1:p},\bar{\mathbf{a}}_{1:p-1})}{q^{N}(\theta,k,\bar{\mathcal{X}}_{1:p},\bar{\mathbf{a}}_{1:p-1})} & =\frac{N^{-p}\overline{\pi}(\theta,\mathcal{X}_{1:p}^{(k)})}{\bar{W}_{p}^{(k)}\mathcal{M}_{1}(\mathcal{X}_{1}^{(b_{1}^{k})})\left(\prod_{n=2}^{p}\bar{W}_{n-1}^{(b_{n-1}^{k})}\mathcal{M}_{n}(\mathcal{X}_{n}^{(b_{n}^{k})}|\mathcal{X}_{n-1}^{(b_{n-1}^{k})})\right)\bar{q}(\theta(i-1)|\theta^{\prime})}\\
 & =\frac{N^{-p}\overline{\gamma}_{p}(\mathcal{X}_{1:p}^{(k)})\overline{p}(\theta)/Z_{p}}{\mathcal{M}_{1}(\mathcal{X}_{1}^{(b_{1}^{k})})\prod_{n=2}^{p}\mathcal{M}_{n}(\mathcal{X}_{n}^{(b_{n}^{k})}|\mathcal{X}_{n-1}^{(b_{n-1}^{k})})\left(\prod_{n=1}^{p}\bar{W}_{n}^{(b_{n}^{k})}\right)\bar{q}(\theta(i-1)|\theta^{\prime})}\\
 & =\frac{\overline{\gamma}_{p}(\mathcal{X}_{1:p}^{(k)})\left(\prod_{n=1}^{p}N^{-1}\left(\sum_{j=1}^{N}w_{n}(\mathcal{X}_{n}^{(j)})\right)\right)\overline{p}(\theta)}{Z\mathcal{M}_{1}(\mathcal{X}_{1}^{(b_{1}^{k})})\left(\prod_{n=2}^{p}\mathcal{M}_{n}(\mathcal{X}_{n}^{(b_{n}^{k})}|\mathcal{X}_{n-1}^{(b_{n-1}^{k})})\right)\left(\prod_{n=1}^{p}w(\mathcal{X}_{n}^{(b_{n}^{k})})\right)\bar{q}(\theta|\theta^{\prime})}\\
 & =\frac{\hat{Z}_{p}}{Z}\times\frac{\overline{p}(\theta)}{\bar{q}(\theta|\theta^{\prime})},
\end{align*}
 where we denote the normalising constant of the posterior in \eqref{eq:target}
as: 
\[
Z=\int_{\Theta}Z_{p}\overline{p}(\theta)d\theta
\]
 Therefore the Metropolis-Hastings procedure to sample from $\bar{\pi}_{p}^{N}$
will be as in Algorithm \ref{fig:stop_within_mcmc}.

Alternatively using similar arguments one we may write 
\[
\overline{\pi}_{p}^{N}(\theta,k,\bar{\mathcal{X}}_{1:p},\bar{\mathbf{a}}_{1:p-1})=\frac{\widehat{Z}_{p}}{Z}\psi_{\theta}(\bar{\mathcal{X}}_{1:p},\bar{\mathbf{a}}_{1:p-1})\bar{W}_{p}^{k}
\]
 Summing over $k$ and using the unbiased property of the SMC algorithm
in Equation \eqref{eq:SMC_unbiased} it follows that $\bar{\pi}_{p}^{N}(\cdot)$
admits $\bar{\pi}(\theta)$ as a marginal, so the proof of part 1.
is complete.

Part 2. is a direct consequence of Theorem 1 in \cite{pseudo-marginal}
and Assumption (A\ref{assump:a5-6}).

\end{proof}

\begin{proof}{[}Proof of Proposition \ref{prop:adap_stop}{]} The
proof is the similar as that of Proposition \ref{prop:stop_within_mcmc}.
For the proof of the first statement of part 1. one repeats the same
arguments as for Proposition \ref{prop:stop_within_mcmc} with difference
being in the inclusion of $\overline{\Lambda}_{\theta}(v)$ for $\bar{\pi}^{N}$
and $\bar{q}^{N}$. For the second statement, to get the marginal
of $\overline{\pi}^{N}$, re-write the target as: 
\[
\overline{\pi}^{N}(\theta,k,v,\bar{\mathcal{X}}_{1:p(v)},\bar{\mathbf{a}}_{1:p(v)-1})=\frac{\widehat{Z}_{p(v)}}{Z}\psi_{\theta}(\bar{\mathcal{X}}_{1:p(v)},\bar{\mathbf{a}}_{1:p(v)-1})\bar{W}_{p(v)}^{k}\overline{\Lambda}_{\theta}(v).
\]
 Let $\overline{\pi}_{p}^{N}(\theta)$ denote the marginal of $\bar{\pi}_{p}^{N}(\cdot)$
obtained in Proposition \ref{prop:stop_within_mcmc}. Using \eqref{eq:cond}
and the conditional independence of $v$ and $\bar{\mathcal{X}}_{1:p(v)},\bar{\mathbf{a}}_{1:p(v)-1}$,
then for the marginal of $\bar{\pi}^{N}(\cdot)$ w.r.t $v$, $\bar{\mathcal{X}}_{1:p(v)},\bar{\mathbf{a}}_{1:p(v)-1}$,
$k$ we have that 
\[
\overline{\pi}^{N}(\theta)=\int_{V}\overline{\pi}_{p(v)}^{N}(\theta)\overline{\Lambda}_{\theta}(v)dv=\overline{\pi}(\theta),
\]
 where the summing over ~$k$ and integrating w.r.t.~$\bar{\mathcal{X}}_{1:p(v)},\bar{\mathbf{a}}_{1:p(v)-1}$
is as in Proposition \ref{prop:stop_within_mcmc}.

For part 2. note that the conditional density given $k$ and $v$
and $\theta$ of $\mathcal{X}_{1:p(v)}^{(k)}$ is 
\[
\frac{\overline{\pi}(\theta,\mathcal{X}_{1:p(v)}^{(k)})\overline{\Lambda}_{\theta}(v)}{\overline{\pi}(\theta)\overline{\Lambda}_{\theta}(v)}=\overline{\pi}(\mathcal{X}_{1:p(v)}^{(k)}|\theta).
\]
 Hence the sequence $\left(\theta(i),\mathcal{X}_{1:p(v)}^{(k)}(i)\right)_{i\geq0}$
satisfies the required property as direct consequence Theorem 1 in
\cite{pseudo-marginal} and Assumption (A\ref{assump:a5-6}). \end{proof}

\end{document}